%% file: ismm20aCameraReady.tex
\documentclass[sigplan,screen]{acmart}

\AtBeginDocument{%
  \providecommand\BibTeX{{%
    \normalfont B\kern-0.5em{\scshape i\kern-0.25em b}\kern-0.8em\TeX}}}

\acmConference[ISMM '20]{Proceedings of the 2020 ACM SIGPLAN International Symposium on Memory Management}{June 16, 2020}{London, UK}
\acmBooktitle{Proceedings of the 2020 ACM SIGPLAN International Symposium on Memory Management (ISMM '20), June 16, 2020, London, UK}

\usepackage{multirow}
\usepackage{gensymb}
\usepackage{graphicx,pifont}
\let\oldding\ding
\renewcommand{\ding}[2][1]{\scalebox{#1}{\oldding{#2}}}

\usepackage{tabu}

\usepackage{color}
\usepackage{balance}
\usepackage{microtype}
\newcommand{\techbold}{\textbf{{DATA\-CON}}}
\newcommand{\tech}{\text{{DATACON}}}
\newcommand{\aclatim}{\text{{31}}}
\newcommand{\perfim}{\text{{27}}}
\newcommand{\enim}{\text{{43}}}
\newcommand{\ao}{\text{{9KB}}}
\newcommand{\changed}[1]{\textcolor{black}{#1}}
\newcommand{\iscachanged}[1]{\textcolor{black}{#1}}
\newcommand{\mcone}[1]{\textcolor{black}{#1}}

\newcommand{\ineq}[1]{\footnotesize$#1$\normalsize}{}
\usepackage[redeflists]{IEEEtrantools}
\usepackage{dblfloatfix}

\begin{document}
\bstctlcite{IEEEexample:BSTcontrol}
\title[Improving Phase Change Memory Performance with Data Content Aware Access]{Improving Phase Change Memory Performance \\with Data Content Aware Access}

\author{Shihao Song}
\affiliation{\institution{Drexel University}\country{USA}}

\author{Anup Das}
\affiliation{\institution{Drexel University}\country{USA}}

\author{Onur Mutlu}
\affiliation{\institution{ETH Z{\"u}rich}\country{Switzerland}}

\author{Nagarajan Kandasamy}
\affiliation{\institution{Drexel University}\country{USA}}
\renewcommand{\shortauthors}{Shihao Song, Anup Das, Onur Mutlu, and Nagarajan Kandasamy}

\begin{abstract}
\input{sections/abstract}
\end{abstract}




\maketitle

\section{Introduction}
\label{sec:introduction}
\input{sections/intro}

\section{Background}
\label{sec:pcm}
\input{sections/pcm}
\section{Data Content Aware Access in PCM}
\label{sec:dat}
\input{sections/dat}

\vspace{-10pt}
\section{Evaluation Methodology}
\label{sec:evaluation}
\input{sections/evaluations}

\vspace{-10pt}
\section{Results and Discussions}
\label{sec:results}

\input{sections/results}
\section{Related Works}
\label{sec:related_works}
\input{sections/related}

\section{Conclusions}
\label{sec:conclusions}
\input{sections/conclusions}

\section*{Acknowledgments}
This work is supported by 1) the National Science Foundation Faculty Early Career Development Award CCF-1942697 (CAREER: Facilitating Dependable Neuromorphic Computing: Vision, Architecture, and Impact on Programmability) and 2) the National Science Foundation Award CCF-1937419 (RTML: Small: Design of System Software to Facilitate Real-Time Neuromorphic Computing).

\balance
\bibliographystyle{IEEEtranSN}
\bibliography{pcm}

\end{document}

%% file: sections/abstract.tex
Phase change memory (PCM) is a scalable non-volatile memory technology that has low access latency (like DRAM) and high capacity (like Flash).
Writing to PCM  incurs significantly higher \emph{latency} and \emph{energy} pe\-nalties compared to reading its content.
A prominent characteristic of PCM's write operation is that its latency and energy are {sensitive} to the data to be written as well as the content that is \emph{overwritten}. 
We observe that overwriting \emph{unknown} memory content can incur significantly higher latency and energy compared to overwriting \emph{known} all-zeros or all-ones content.
This is because all-zeros or all-ones content is overwritten by programming the PCM cells \emph{only} in one direction, i.e., using either SET or RESET operations, not both.

In this paper, we propose \emph{data content aware PCM writes} (\techbold), a new mechanism that reduces the latency and energy of PCM writes by redirecting these requests to overwrite memory locations containing all-zeros or all-ones.	
\tech~operates in three steps.
First, it estimates how much a PCM write access would benefit from overwriting known content (e.g., all-zeros, or all-ones) by comprehensively considering the number of set bits in the data to be written, and the energy-latency trade-offs for SET and RESET operations in PCM.
Second, it translates the write address to a physical address within memory that contains the best type of content to overwrite, and records this translation in a table for future accesses. We exploit data access locality in workloads to minimize the address translation overhead.
{
Third, it re-initializes \emph{unused} memory locations with known all-zeros or all-ones content in a manner that does not interfere with regular read and write accesses.
}
\tech~overwrites unknown content only when it is absolutely necessary to do so. 
{
We evaluate \tech~with workloads from state-of-the-art machine learning applications, SPEC CPU2017, and NAS Parallel Benchmarks.
}
Results demonstrate that \tech~improves the effective access latency by \aclatim\%, overall system performance by \perfim\%,  and total memory system energy consumption by \enim\% compared to the best of performance-oriented state-of-the-art techniques.

%% file: sections/intro.tex
DRAM has been the technology choice for implementing main memory due to its relatively low latency and low cost. 
{
However, DRAM is a fundamental performance and energy bottleneck in almost all computing systems~\cite{wilkes2001memory,mutlu2015research,mutlu2013memory,mandelman2002challenges}, and it is experiencing significant technology scaling challenges~\cite{mutlu2013memory,kang2014co,mutlu2017rowhammer,mutlu2019rowhammer,kim2014flipping,frigo2020trrespass}.
}
Recently, DRAM-compati\-ble, yet more technology scalable alternative me\-mory technologies are being explored: phase-change memory (PCM)~\cite{LeeISCA2009,lee2010phase,LeeMicro10,qureshi2010improving,QureshiISCA09,qureshi2011pay,kannan2016energy,ZhangASPLOS15,ferreira2010increasing,bock2011analyzing,jiang2012fpb,zhou2012writeback,DuISCA13,zhou2013writeback,jiang2013hardware,yue2013accelerating,ZhouISCA09,yamada2009100,sebastian2017temporal,burr2008overview,zhang2016mellow,wang2015exploit,liu2018crash,atwood2018pcm}, spin-transfer torque magnetic RAM (STT-MRAM)~\cite{kultursay2013evaluating}, 
{
metal-oxide resistive RAM (Re\-RAM)~\cite{akinaga2010resistive,wong2012metal,mallik2017design},
}
conductive bridging RAM (CB\-RAM)~\cite{kund2005conductive}, ferro-electric RAM (FeRAM)~\cite{bondurant1990ferroelectronic}, etc.
Of the various emerging memory technologies, PCM is the most mature~\cite{lai2008non,wong2010phase} due to its CMOS compatibility~\cite{jo2008cmos},  deep material understanding~\cite{wuttig2007phase}, and availability of high-capacity working prototypes~\cite{villa201045nm,ahn2005highly,oh2006full,lee2007highly,im2008unified,kang2011pram}.

However, PCM has finite write endurance, and its access latency and energy are higher than those of DRAM~\cite{LeeISCA2009,QureshiISCA09,joo2010energy,chen2011rpram,pirovano2004reliability,SeongSecurityISCA2010,chen2009endurance,jiang2014mitigating}.
One key characteristic of PCM is the \emph{asymmetry} in read and write latencies \cite{yoon2014efficient,lung2016double,LeeISCA2009}.

This is because the SET operation ($0\rightarrow 1$ programming) in PCM requires longer latency than the RESET operation ($1\rightarrow 0$ programming).
Due to memory bank interleaving, multiple requests can be served in parallel using different PCM banks \cite{lung2016double}.
Unfortunately, when {two} requests go to the \emph{same bank}, they have to be served \emph{serially}. This is known as \emph{bank conflict}.
{
Slow writes in PCM increase bank conflict latencies, degrading system performance~\cite{song2019enabling}.
}

Modern computing systems are embracing hybrid memories comprising of DRAM and PCM to combine the best properties of both memory technologies, achieving low latency, high reliability, and high density.
3D XPoint memory is one example of hybrid memory with DRAM and PCM connected to separate DIMMs~\cite{bourzac2017has}.
The IBM POWER 9 processor~\cite{sadasivam2017ibm} is another example, where 
embedded DRAM (eDRAM) is used as a write cache to PCM main memory.
Many recent works mitigate the impact of slow and costly PCM writes in hybrid memories by migrating write-intensive pages from PCM to DRAM~\cite{QureshiISCA09,dhiman2009pdram,liu2018crash,li2017utility,zhao2014firm,yoon2012row,meza2012enabling,hsieh2012double,ham2013disintegrated,hwang2013hmmsched,bock2016concurrent,bock2015understanding,bock2015characterizing,bock2015hmmsim,bock2014concurrent,jia2018vail,meza2013case,ren2015thynvm}.
Unfortunately, \emph{none} of these approaches \emph{fundamentally} reduce write latency at its source and PCM latency continues to be a performance bottleneck, especially for emerging data-intensive applications, \changed{where high volumes of data are accessed frequently.}


\changed{One simple approach to reducing PCM latency is to use \emph{only} RESET operations during all PCM writes~\cite{lam2010block,QureshiISCA12}.
These schemes require the PCM cells in a memory location to be SET, prior to a write.}
Unfortunately, if a PCM write induces many RESET operations, the energy consumption of such schemes can be higher than simply overwriting the existing memory content as in a baseline system~\cite{LeeISCA2009}. This is because a RESET operation in PCM requires higher energy than a SET operation.
We observe that for some workloads, the energy overhead of such a RESET-only scheme can be up to 30\% over Baseline (Section \ref{sec:perf}).
Recent works also propose to reduce the energy overhead of PCM write operations by selectively programming the PCM cells~\cite{cho2009flip,chen2011energy,yang2007low,mirhoseini2012coding,fang2012softpcm,jacobvitz2013coset}. This is achieved by analyzing the data to be written (referred to as \textbf{write data}) and the content of the memory location that is overwritten (referred to as \textbf{overwritten content}). However, such schemes increase the effective write latency by over 15\% due to the additional read request required to find out the overwritten content (Section \ref{sec:perf}).

\textbf{Our goal} is to design a mechanism to \changed{reduce the latency and energy of PCM writes.}
{
We achieve this goal by exploiting the following three major observations in this paper.
}

\begin{figure}[h!]
	\centering
	\vspace{-5pt}
	\centerline{\includegraphics[width=0.99\columnwidth]{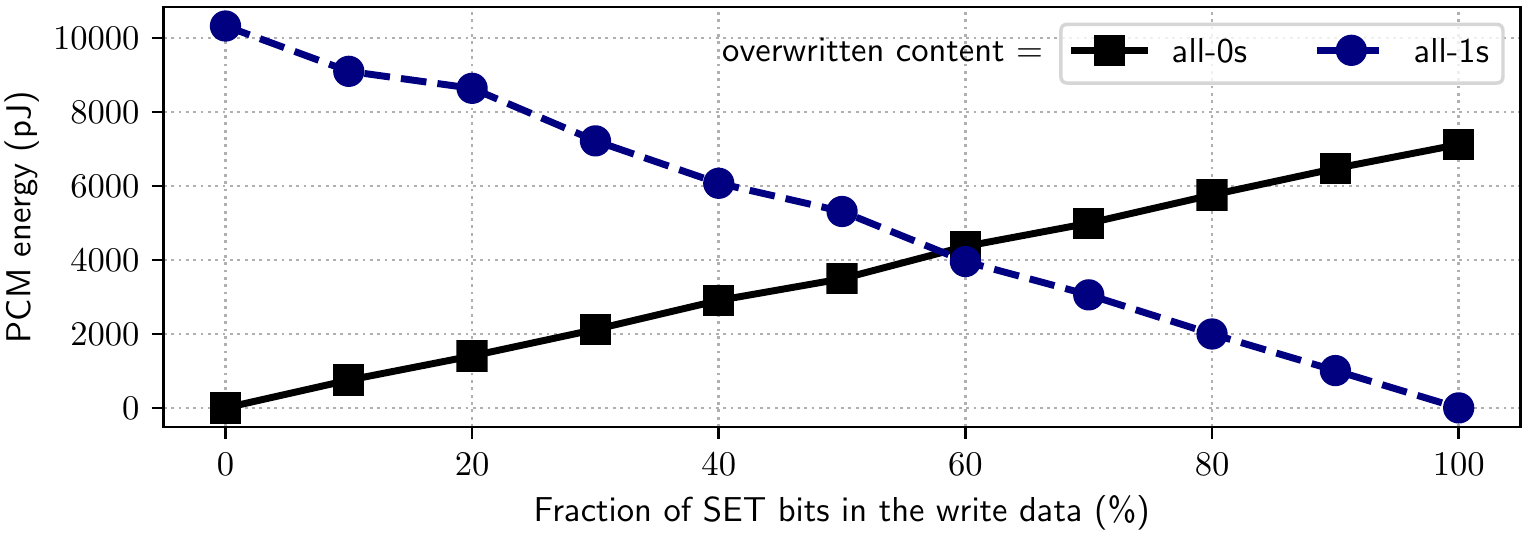}}
	\vspace{-5pt}
	\caption{{Energy of a PCM write for increasing fraction of SET bits in the write data when the overwritten content is all-0s and all-1s.}}
	\label{fig:set_reset_energy}
	\vspace{-10pt}
\end{figure}

\textbf{Observation 1:} 
\textit{
We observe that the energy consumed to serve a PCM write depends on both the write data and the overwritten content. 
}
Figure \ref{fig:set_reset_energy} plots the energy consumption of a PCM write as the fraction of SET bits in the write data is increased from 0\% to 100\% with the overwritten content (i.e., the initial content) set to all-0s and all-1s (see Section \ref{sec:evaluation} for our simulation parameters).
We observe that if the fraction of SET bits in the write data is less than 60\%, overwriting all-0s is more energy efficient than overwriting all-1s. Conversely, if the fraction of SET bits in the write data is more than 60\%, overwriting all-1s is more energy efficient. 
Therefore, \emph{PreSET}~\cite{QureshiISCA12}, a mechanism that initializes a memory location to all-1s before overwriting it, is beneficial only in the latter scenario, where the fraction of SET bits in the write data is more than 60\%.

\begin{figure}[h!]
	\centering
	\vspace{-5pt}
	\centerline{\includegraphics[width=0.99\columnwidth]{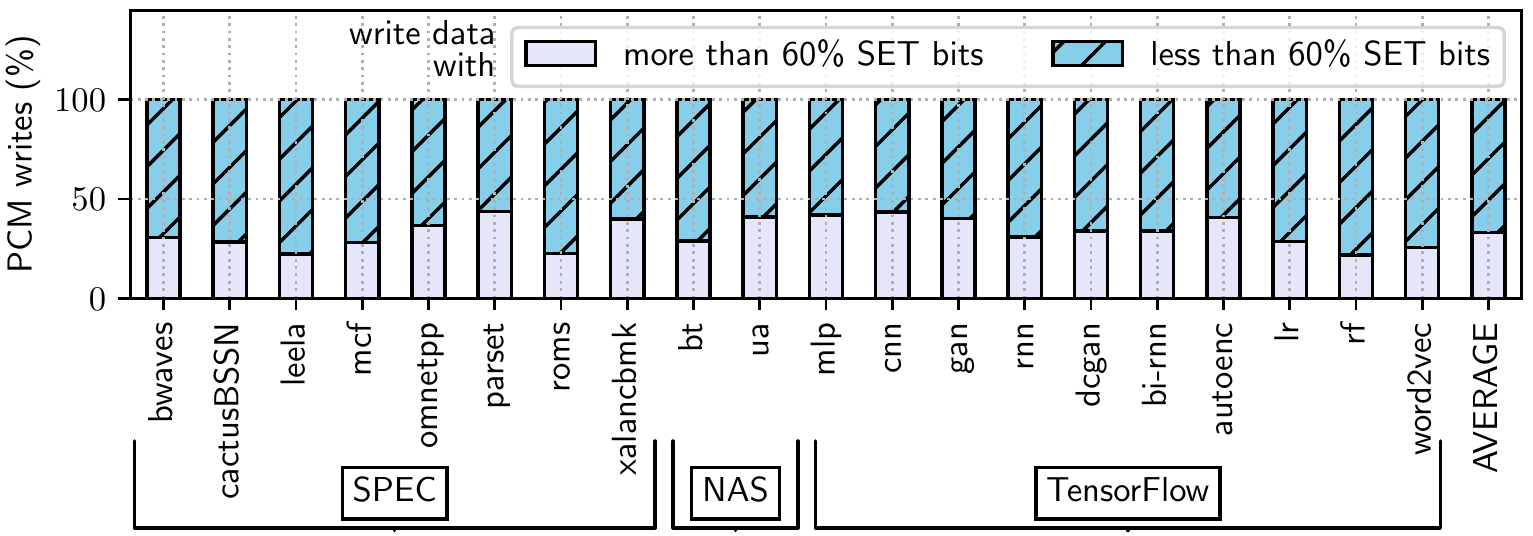}}
	\vspace{-5pt}
	\caption{{Total number of PCM writes distributed into write accesses having more than 60\% and less than 60\% SET bits in their write data for each of the evaluated workloads.}}
	\vspace{-5pt}
	\label{fig:set_reset_distribution}
\end{figure}

\textbf{Observation 2:} 
\textit{
We observe that the number of SET bits in the write data is workload dependent. 
}
Figure \ref{fig:set_reset_distribution} plots the total number of PCM writes distributed into write accesses 
having more than 60\% and less than 60\% SET bits in their write data for each of the evaluated workloads (Section \ref{sec:evaluation}).
We observe that 
on average, only 33\% of PCM writes in these workloads have more than 60\% SET bits in their write data.
Therefore, PreSET can bring energy benefit only for a small fraction of PCM write accesses, leaving behind a significant opportunity for energy improvement. 
Exploiting this opportunity will not only reduce energy but also improve performance by enabling more concurrency.

We propose \emph{data content aware PCM writes} (\techbold), a new mechanism to reduce the latency and energy of PCM writes by exploiting these two observations.
The \textbf{key idea} is to estimate how much a PCM write access would benefit from overwriting \emph{known} all-0s or all-1s content by comprehensively considering the fraction of SET bits in the write data and the energy-latency trade-offs for SET and RESET operations in PCM. 
Based on this estimate, \tech~selects a memory location to redirect the write request to, containing the \emph{best} overwritten content.
%
%
%

The operation of \tech~is straightforward. 
Wh\-en a cache block is evicted from DRAM, the memory controller redirects the write request to a new physical address within memory, which contains the best overwritten content for the write data. 
When overwriting this content, the latency and energy of the write operation reduces due to programming the PCM cells \emph{only} in one direction, i.e., using only SET or RESET operations, not both. This improves system performance. 
The memory controller maintains a table to record all such address translations.
As memory locations are continuously re-mapped,
\tech~periodically invalidates entries from the table and re-initializes \emph{unused} memory locations with all-zeros or all-ones so that future write requests can be serviced using them. 
Section \ref{sec:dat} describes the implementation of \tech.
To limit the size of \tech{}'s address translation table, we exploit the following observation.

\iscachanged{\textbf{Observation 3:}  
We observe that a modern PCM bank is implemented as a collection of \emph{partitions} that operate mostly \emph{independently} while \emph{sharing} a few global peripheral structures, which include the sense amplifiers (to read) and the write drivers (to write) \cite{song2019enabling}. 
{Many workloads tend to frequently access rows that belong to the \emph{same} memory partition in a bank.} 
We refer to this form of spatial locality where rows from the same partition are read from and written to frequently as \emph{Partition Level Spatial Locality (PLSL)}.}

\tech{} exploits the partition-level spatial locality in a workload to \emph{cache} only a part of the address translation table inside the memory controller, while the full address translation table is stored in memory. 

We evaluate \tech{} with workloads from state-of-the-art machine learning applications \cite{tfex}, SPEC~CPU2017~\cite{bucek2018spec}, and NAS Parallel Benchmarks \cite{bailey1991parallel}. Our results show that \tech{} improves performance and energy efficiency over state-of-the-art techniques that reduce latency and energy of PCM writes. For multi-core workloads, \tech~improves the effective access latency by \aclatim\%, overall system performance by \perfim\%, and total system energy consumption by \enim\% with a hardware cost of only \ao~ for a PCM memory of 128GB capacity. As DATACON reduces the latency and energy of a PCM write access at its source, \tech{} can be combined with other mechanisms that aim to reduce the \emph{number} of write accesses to PCM~\cite{dhiman2009pdram,li2017utility,zhao2014firm,yoon2012row,meza2012enabling,hsieh2012double,ham2013disintegrated,hwang2013hmmsched,bock2016concurrent,bock2015understanding,bock2015characterizing,bock2015hmmsim,bock2014concurrent,xia2014dwc,hu2013write,huang2011register,zhang2016mellow}.

We make the following \textbf{contributions}.

\vspace{-5pt}

\begin{itemize}
	\item We propose an efficient technique, \tech, wh\-ich lowers the PCM write latency and energy by overwriting the \emph{best} content on a write access. DA\-TACON overwrites unknown content only when it is absolutely necessary to do so. DATACON maintains a table for address translations and methodically re-initializes \emph{unused} memory locations to service future write requests.
	\item \iscachanged{We observe that many workloads exhibit a form of locality where rows from the \emph{same} memory partition are frequently read from and written to frequently. We refer to this locality as \emph{Partition Level Spatial Locality (PLSL)}. \tech~exploits this locality to minimize the overhead of its address translation table by \emph{caching} address translations of only the frequently accessed PCM partitions.}
	\item We comprehensively evaluate the performance and energy efficiency of \tech. Our results show that \tech~significant\-ly improves performance and energy efficiency acr\-oss a wide variety of workloads.
\end{itemize}

%% file: sections/pcm.tex
\subsection{PCM Operation}
\begin{figure}[h!]
	\begin{center}
		\vspace{-10pt}
		\includegraphics[width=0.99\columnwidth]{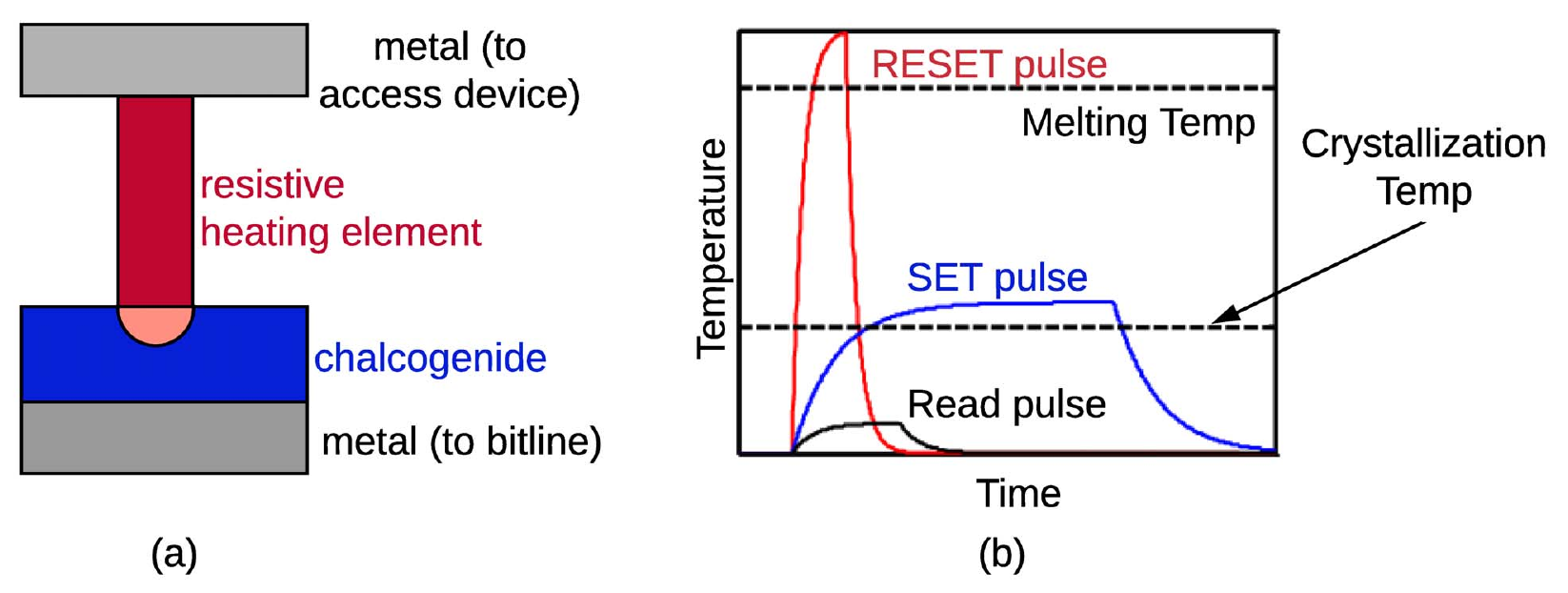}
		\vspace{-10pt}
		\caption{(a) A phase change memory (PCM) cell and (b) current needed to SET, RESET, and read a PCM cell.}
		\vspace{-10pt}
		\label{fig:pcm_memory_cell_integration}
		\vspace{-10pt}
	\end{center}
\end{figure}

Figure \ref{fig:pcm_memory_cell_integration}(a) illustrates how a chalcogenide semiconductor alloy is used to build a PCM cell.
{
The amorphous phase (logic `0') in this alloy has higher resistance than the crystalline phase (logic `1').
}
Ge${}_2$Sb${}_2$Te${}_5$ (GST) is the most commonly used alloy for PCM~\cite{ovshinsky1968reversible,lai2003current,bez2009chalcogenide,burr2010phase,salinga2018monatomic,wang2018scandium} due to its high amorphous-to-crystalline resistance ratio, fast switching between phases, and high endurance.
However, other chalcogenide alloys are also explored due to their better data retention properties~\cite{morikawa2007doped,lin2007nano,zhang2007te}.
Phase changes in a PCM cell are induced by injecting current into the resistor-chalcogenide junction and heating the chalcogenide alloy. 

{
Figure \ref{fig:pcm_memory_cell_integration} (b) shows the different current profiles needed to program and read in a PCM device~\cite{secco2018flux}.
}
To RESET a PCM cell, a high power pulse of short duration is applied and quickly terminated. This first raises the temperature of the chalcogenide alloy to 650$\degree$C, above its melting point. The melted alloy subsequently cools extremely quickly, locking into an amorphous phase.
To SET a PCM cell, the chalcogenide alloy is heated above its crystallization temperature, but below its melting point for a sufficient amount of time. 
Finally, to read the content (i.e., know the phase) of a PCM cell, a small electrical pulse is applied that is sufficiently low so as not to induce phase change in the PCM cell. 

\subsection{PCM Organization}
To ensure reasonable memory read and write latencies at high density, a PCM chip, like the DRAM, is organized hierarchically~\cite{villa2018pcm}.
For example, a 128GB PCM can have 2 {channels}, 1 rank/channel and 8 banks/rank. A bank can have {64} {partitions}~\cite{song2019enabling}, which are similar to subarrays in DRAM~\cite{KimISCA12,chang2017understanding}. 
Within each partition, 
the PCM cells are organized as an array. A column of PCM cells is called a 
\textit{bitline}.
A row of PCM cells is called a \emph{wordline}. 
Each PCM bank can have 128 peripheral circuits to read and program 128 bits in the array in parallel.

\subsection{PCM Timings}
We briefly introduce the commands and timing parameters of PCM to understand \tech{}. 
Without loss of generality, we describe them for the 28nm PCM design from Micron \cite{villa2018pcm}.
To serve a memory request that accesses data at a particular row and column address within a bank, a memory controller issues \emph{three} commands to the bank~\cite{song2019enabling}.
\begin{itemize}
	\item \texttt{\underline{ACTIVATE}:} activate the wordline and enable the peripheral circuit for the memory cells to be accessed.
	\item \texttt{\underline{READ}/\underline{WRITE}:} drive read or write current through the cell. After this command executes, the data stored in the cell is available at the output terminal of peripheral circuit, or the write data is programmed to the cell.
	\item \texttt{\underline{PRECHARGE}:} deactivate the wordline and bitline, and prepare the bank for the next access. 
\end{itemize}

\begin{figure}[h!]
 	\centering
    \vspace{-5pt}
 	\centerline{\includegraphics[width=0.99\columnwidth]{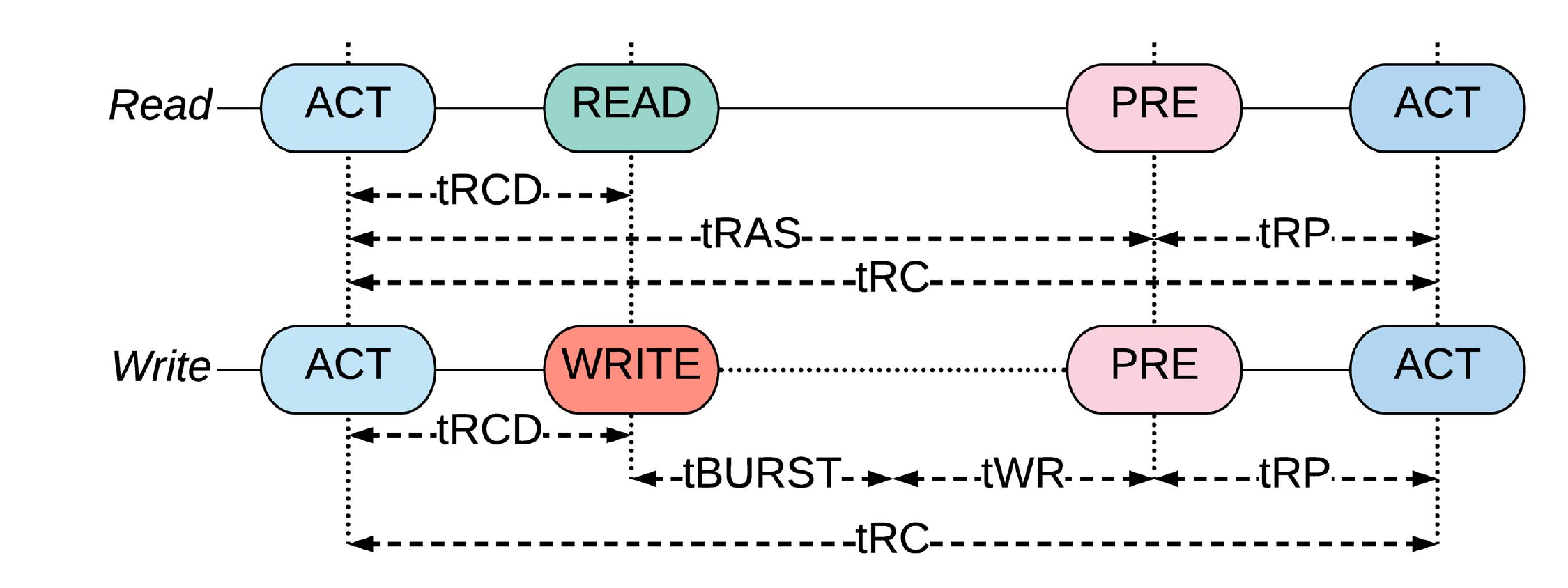}}
 	\vspace{-10pt}
 	\caption{PCM timings for serving read and write requests.}
    \vspace{-20pt}
 	\label{fig:memory_timings}
\end{figure}

Figure~\ref{fig:memory_timings} shows the different memory timing parameters used in \tech{}.
Table~\ref{tab:min_max_latency} provides the 
typical values 
of these timing parameters
based on Micron's 28nm PCM design~\cite{villa2018pcm}.\footnote{{We expect these values to be similar for other PCM designs.}}
Table~\ref{tab:min_max_latency} also reports the timing parameters used in \tech{} to service PCM reads and writes. \tech{} uses the SET and RESET timing parameters when overwriting all-0s and all-1s content, respectively. \tech{} uses the long-latency baseline write timing parameters only when overwriting unknown content. 
Our technique improves performance by minimizing the number of times unknown content is overwritten.

\begin{figure*}[t!]
	\centering
	\centerline{\includegraphics[width=1.7\columnwidth]{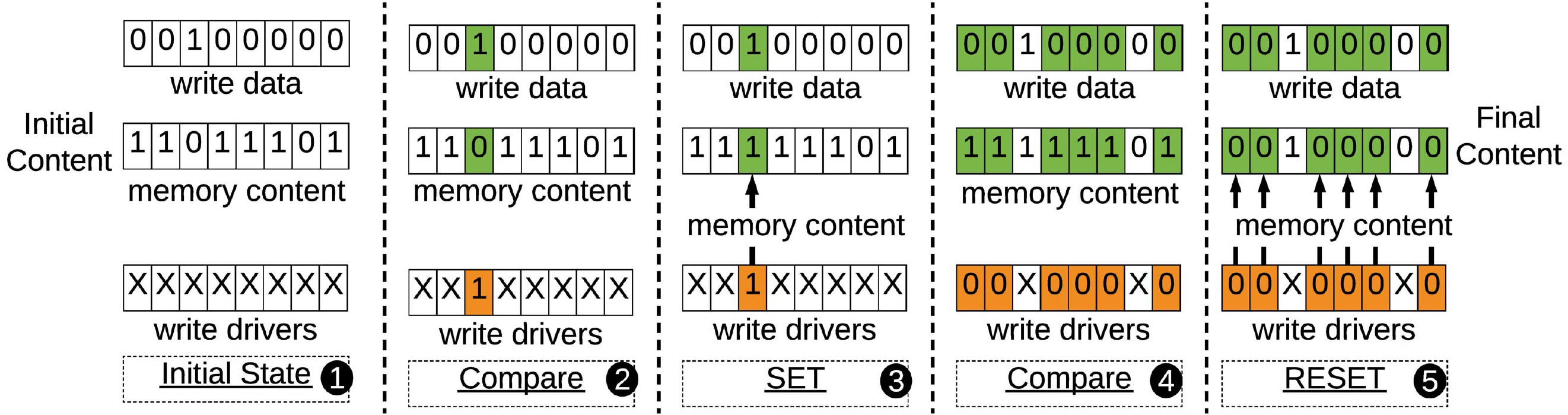}}
	\vspace{-10pt}
	\caption{Servicing a PCM write request in a baseline system~\cite{LeeISCA2009}.}
	\vspace{-10pt}
	\label{fig:pcm_write}
\end{figure*}

\begin{figure*}[t!]
	\centering
	\centerline{\includegraphics[width=2.01\columnwidth]{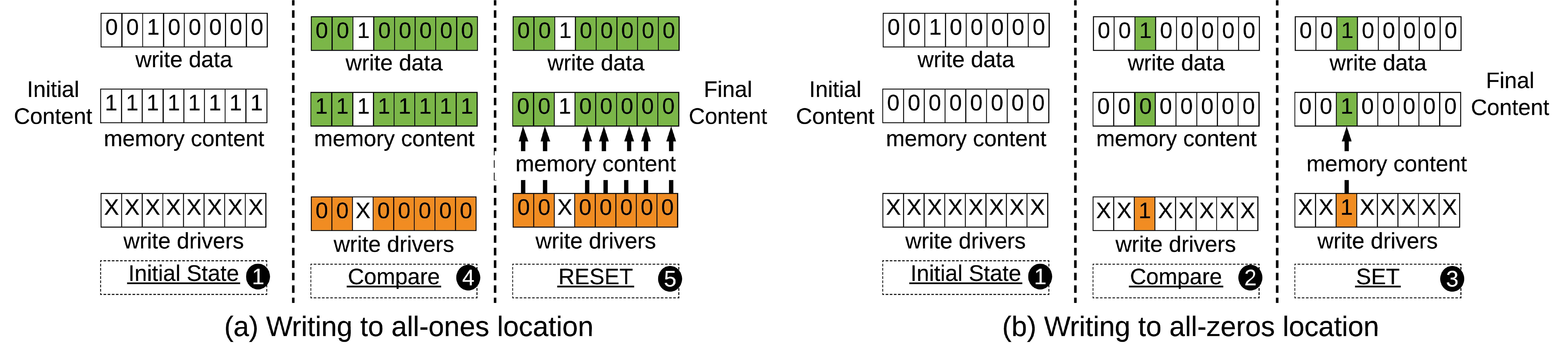}}
	\vspace{-10pt}
	\caption{Servicing a PCM write by overwriting all-1s (a) and all-0s (b).}
	\vspace{-10pt}
	\label{fig:pcm_write_0_1}
\end{figure*}

\section{Performance and Energy Improvements in \tech{}}
\label{sec:motivation}
\subsection{Quantifying Performance Improvement}\label{sec:quantifying_perf}
We present some motivating examples to provide intuition for the performance improvement in DATACON.
Figure \ref{fig:pcm_write} illustrates an 8-bit write operation in PCM. 
The write data `00100000' is to be programmed to a memory location, whose current content is `11011101' (\ding[1]{182}).
This content is unknown at the time of the write operation.
The following four steps are performed in sequence to service the write request in a Baseline PCM chip~\cite{LeeISCA2009}.
\begin{itemize}
	\item[1:] The write data is compared with the overwritten content to select the PCM cells that need to be SET. \changed{In this example, only the third bit from the left needs to be SET. This is achieved by setting bit 3 in the write driver and masking others (\ding[1]{183}).}
	\item[2:]  The selected PCM cell is SET using a long-latency SET pulse generated by the write driver (\ding[1]{184}). The memory content changes to `11111101'. 
	\item[3:] The write data is compared with the new memory content to select the PCM cells that need to be RESET. This is achieved by programming the corresponding bits in the write driver and masking others (\ding[1]{185}).
	\item[4:] The selected PCM cells are RESET by the write driver using short-latency RESET pulses (\ding[1]{186}).
\end{itemize}
\changed{We observe that after executing these four steps the memory content is overwritten with the write data.}

\begin{table}[h!]
\setlength{\tabcolsep}{5pt}
\renewcommand{\arraystretch}{1.0}
\centering
\vspace{-5pt}
\caption{{PCM timing parameters based on \cite{villa2018pcm}.}}
\vspace{-5pt}
\label{tab:min_max_latency}
{\fontsize{8}{10}\selectfont
\begin{tabu}{c c c c c c}
    \tabucline[2pt]{-}
    \multicolumn{6}{c}{\textcolor{red}{\emph{\textbf{Baseline Timing Parameters}}}}\\
    \hline
    & \textbf{tRCD} & \textbf{tRAS} & \textbf{tRP} & \textbf{tRC} & \\\cline{2-6}
    Read & 3.75ns & 55.25ns & 1ns & 56.25ns &  \\
    \hline
    & \textbf{tRCD} & \textbf{tBURST} & \textbf{tWR} & \textbf{tRP} & \textbf{tRC}\\\cline{2-6}
     Write & 75ns & 15ns & 190ns & 1ns & 209.75ns \\
     \hline
     \multicolumn{6}{c}{\textcolor{red}{\emph{\textbf{\tech{} Timing Parameters}}}}\\
     \hline
     & \textbf{tRCD} & \textbf{tRAS} & \textbf{tRP} & \textbf{tRC} & \\\cline{2-6}
    Read & 3.75ns & 55.25ns & 1ns & 56.25ns &  \\
    \hline
    & \textbf{tRCD} & \textbf{tBURST} & \textbf{tWR} & \textbf{tRP} & \textbf{tRC}\\\cline{2-6}
     SET (all-0s) & 3.75ns & 15ns & 150ns & 1ns & 169.75ns \\
     RESET (all-1s) & 3.75ns & 15ns & 40ns & 1ns & 59.75ns \\
     Write (unknown) & 75ns & 15ns & 190ns & 1ns & 209.75ns \\
    \tabucline[2pt]{-}
\end{tabu}
}
\vspace{-10pt}
\end{table}

Figure \ref{fig:pcm_write_0_1}(a) and (b) illustrate, respectively, two scenarios -- one where the write data is overwritten on all-1s (as in PreSET), and another one where it is overwritten on all-0s.
We observe that for overwriting all-1s, step 3 (\emph{compare}) \ding[1]{185} and step 4 (\emph{RESET}) \ding[1]{186} are sufficient, while for overwriting all-0s, step 1 (\emph{compare}) \ding[1]{183} and step 2 (\emph{SET}) \ding[1]{184} are sufficient. Thus, in both these scenarios only two out of the conventional four steps are sufficient for servicing a PCM write, which improves both performance and energy consumption.

Using the timing parameters from Table \ref{tab:min_max_latency}, \tech{} leads to 71.5\% and 19\% lower PCM write latency than Baseline, when using the RESET and SET timings, respectively.
We conclude that overwriting all-1s (i.e., using only RESET operations) has a definitive performance advantage versus overwriting all-0s. In Section \ref{sec:results}, we demonstrate an average \perfim\% performance improvement with \tech{} over PreSET~\cite{QureshiISCA12} for all our evaluated workloads.

\subsection{Quantifying Energy Improvement}
Table~\ref{tab:pcm_write} reports the energy consumption of the three write scenarios of Figures \ref{fig:pcm_write} \& \ref{fig:pcm_write_0_1}, i.e., overwriting the unknown content (`11011101'), overwriting all-0s (`00000000'), and overwriting all-1s (`11111111') for the same write data (`00100000').

The total energy consumption includes (1) the energy to prepare the memory content (\emph{preparation energy}), i.e., to program the memory content to all-0s or all-1s prior to servicing the write request, and (2) the energy to service the write request (\emph{service energy}) by overwriting the memory content with the write data. 
The preparation step is \emph{not} present when overwriting \emph{unknown} content as in a Baseline system~\cite{LeeISCA2009}.
Preparing a memory content with all-1s (required by the PreSET technique of Qureshi et al. \cite{QureshiISCA12}) requires lower energy than preparing the location with all-0s (27pJ for the former vs. 115.2pJ for the latter). This is because only 2 SET operations are required in the former scenario, while 6 RESET operations are required in the latter scenario to prepare the data content for overwriting. 
Conversely, the service energy to overwrite all-1s is 10X higher than that to overwrite all-0s (column 4).
This is because to overwrite all-1s, we need 7 RESET operations (see Figure \ref{fig:pcm_write_0_1}(a)), which consume 134.4pJ of energy vs. 13.5pJ to overwrite all-0s, for which we only need 1 SET operation (see Figure \ref{fig:pcm_write_0_1}(b)).
We conclude that overwriting all-0s has a definitive energy advantage (20\% lower energy than overwriting all-1s, and 11\% lower energy than overwriting unknown content). 
In Section \ref{sec:results}, we demonstrate an average \enim\% energy improvement with \tech{} over PreSET~\cite{QureshiISCA12} for all our evaluated workloads.

\begin{table}[h!]
	\centering
	\vspace{-5pt}
	\caption{Energy consumption of the three overwrite scenarios of Figures \ref{fig:pcm_write} and \ref{fig:pcm_write_0_1}.}
	\label{tab:pcm_write}
	\vspace{-5pt}
	\setlength{\tabcolsep}{2pt}
	{\fontsize{8}{10}\selectfont
		\begin{tabular}{|c|c||c|c|c|}
			\hline
			\multirow{2}{*}{\textbf{Write Data}} & \multirow{2}{*}{\textbf{Overwritten Content}} & \multicolumn{3}{|c|}{\textbf{Total Energy (pJ)}}\\
			\cline{3-5}
			& & \textbf{Preparation} & \textbf{Service} & \textbf{Total} \\
			\hline
			\hline
			\multirow{3}{*}{00100000} & 11011101 (unknown) & 0 & 144.7 & \textbf{144.7}\\
			& 00000000 & 115.2 & 13.5 & \textbf{128.7}\\
			& 11111111 (PreSET~\cite{QureshiISCA12}) & 27 & 134.4 & \textbf{161.4}\\
			\hline
	\end{tabular}}
\end{table}


This example clearly demonstrates the latency-energy trade-offs in overwriting all-0s and all-1s content in PCM.
Specifically, SET operations (to overwrite all-0s) are good for energy, while RESET operations (to overwrite all-1s) are good for performance. 
The exact energy and performance trade-off depends on the number of bits that are to be SET or RESET when overwriting content.
Based on this motivating example, we now present our mechanism, \tech, \emph{data content aware PCM writes}.

%% file: sections/dat.tex
\begin{figure}[h!]
	\centering
	\vspace{-5pt}
	\centerline{\includegraphics[width=0.99\columnwidth]{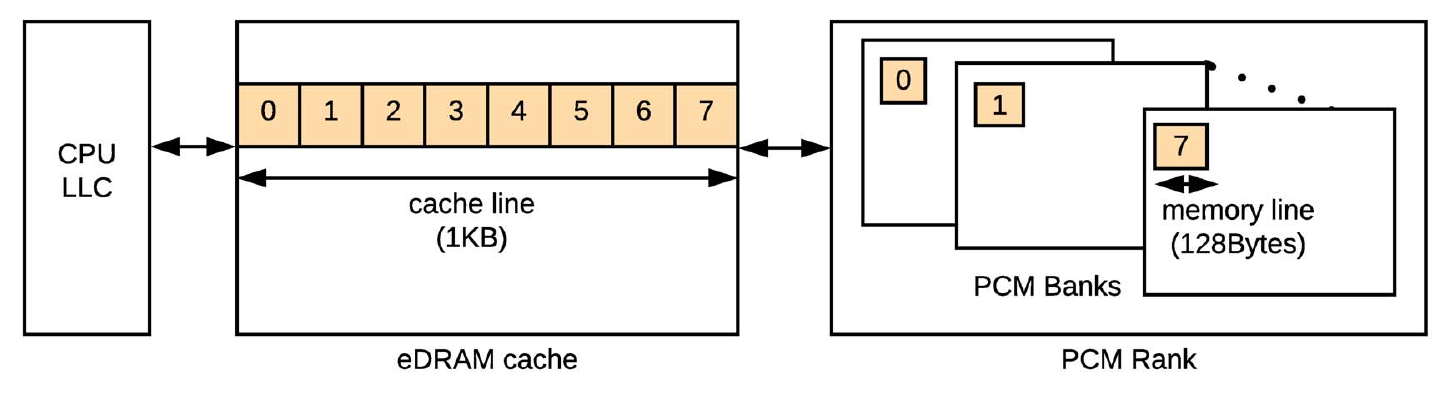}}
	\vspace{-10pt}
	\caption{Mapping of an eDRAM cache line to 8 memory lines in a PCM rank, on our Baseline system.}
	\vspace{-5pt}
	\label{fig:mapping}
\end{figure}

We describe \tech{} in the context of DRAM-PCM hybrid memory, where embedded DRAM (eDRAM) is used as a write cache to PCM main memory.\footnote{Even though we use embedded DRAM as cache to PCM in our implementation and evaluations, \tech{} is applicable to any type of hybrid memory or standalone PCM memory.}
Figure \ref{fig:mapping} illustrates how an eDRAM cache line is mapped to 8 memory lines in a PCM rank. For completeness, we also show the shared last level cache (LLC) of the host CPU.
For a {read miss} in LLC, the memory controller first checks to see if the read address is cached in eDRAM. If cached, the read data is sent from eDRAM to LLC. Otherwise, the memory controller issues a read request to PCM.
The PCM read data is sent back to the requesting CPU, 
{bypassing} eDRAM. 
For a {write miss} in LLC, the memory controller checks to see if the write address is cached in eDRAM. If cached, the cache line is updated with the write data. Otherwise, a read request is issued to PCM. The PCM data is cached in eDRAM and also sent back to the requesting CPU.
In doing so, the least-recently used eDRAM cache line is evicted, which generates a write request to PCM.
These long-latency write requests to PCM are the requests \tech{} optimizes to improve both performance and energy consumption.

We now describe \tech{}'s address translation mechanism, starting with a high-level overview.

\subsection{High-level Overview}
\begin{figure}[h!]
	\centering
	\centerline{\includegraphics[width=0.99\columnwidth]{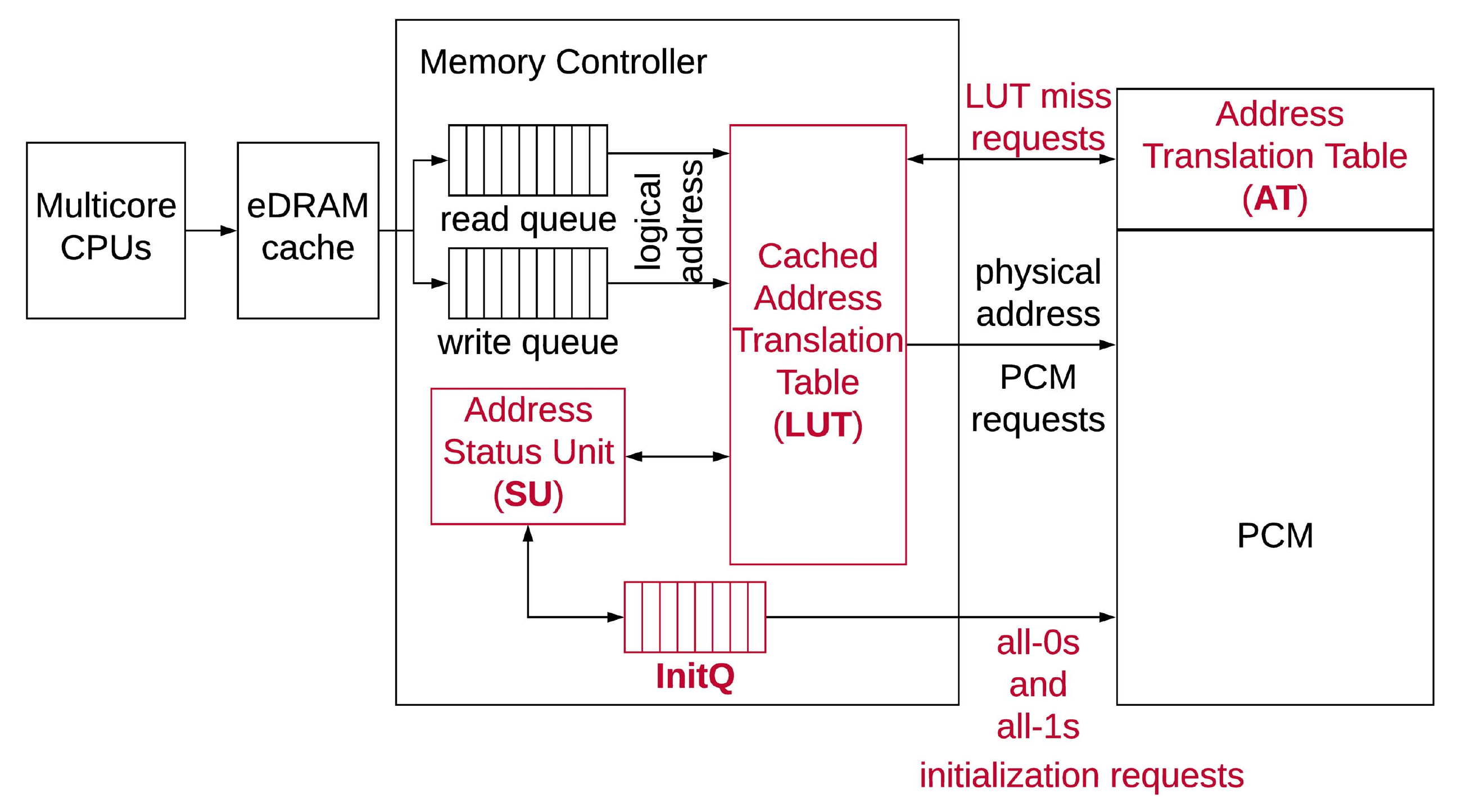}}
	\vspace{-10pt}
	\caption{\tech{}'s address translation mechanism.}
	\vspace{-10pt}
	\label{fig:att}
\end{figure}

Figure \ref{fig:att} shows our system architecture.
For a write request from the \textit{write queue}, 
DATACON performs three operations. 
First, it selects the content to overwrite (all-0s or all-1s) based on the write data and the energy-latency trade-off for overwriting all-0s and all-1s (Section \ref{sec:prioritization}).
Second, it selects a physical address containing the appropriate content to overwrite for servicing the write request and records this logical-to-physical address translation in a table (Section \ref{sec:translation}).
Third, it schedules the write request to PCM.

For a read request from the \textit{read queue}, \tech{} performs two operations. First, it performs the logical-to-physical address translation for the read address using the table (Section \ref{sec:translation}). Second, it schedules the read request to PCM.
\tech{} also methodically re-initializes \emph{unused} memory locations with all-0s or all-1s content in a manner that minimizes interference with PCM accesses (Section \ref{sec:scheduling}).

\vspace{-5pt}

\subsection{Detailed Design of \tech}
Figure \ref{fig:att} shows the detailed design of the memory controller to support \tech{}'s address translation mechanism.
DA\-TACON adds four new components to the baseline memory controller design, which are highlighted in the figure. We describe these components in detail.

The \textit{first} component is the \emph{address translation table} (\textbf{AT}). \tech{} uses this table to record logical-to-physical address translations, which are needed to redirect write requests to the best overwritten content in PCM. AT requires one translation entry for every possible eDRAM cache line (e.g., of 1KB size). With 8GB/1KB = 8,388,608 cache lines per 8GB rank (see our simulation parameters in Table \ref{tab:config}), \tech{} requires 23MB of storage per rank for AT. We use a PCM partition inside the rank to store the AT.

To translate a logical address on demand, an access to the PCM partition storing the AT is required, which can increase the address translation latency. 
To address this, \tech{} uses a second component, 
the \emph{lookup table} (\textbf{LUT}), which reduces the latency of address translation by caching
recently-used address translation information in the memory controller.
We exploit the data access locality of a workload to cache the address translation information of a small number of recently-accessed cache lines in the LUT.
Under this new caching mechanism, the memory
controller categorizes each PCM request from eDRAM into one of two cases: 1) \emph{LUT hit:}, i.e., the translation of the logical address is in LUT, and ii) \emph{LUT miss:}, i.e., the translation of the logical address is not in LUT.
In the first case, the logical address is translated using information from LUT and the request is scheduled to PCM.
In the second case, the memory controller first issues a PCM request to read the address translation table (AT), caches the required AT entry in the LUT, and then translates the logical address using this entry. To make space inside LUT, address translation of the least recently used (LRU) cache line is evicted from LUT. If the evicted content is dirty, a PCM request is generated to write the updated content into the PCM partition storing the AT. In Section~\ref{sec:caching}, we evaluate the size of LUT and 
find that storing the address translation information of only two PCM partitions in LUT (requiring 32KB of storage) is sufficient for the evaluated workloads.

The \textit{third} component is the \textit{address status unit} (\textbf{SU}). DATA\-CON uses this unit to select a physical address for the logical address of a write request.
Internally, SU implements two queues: 1) a 32-entry \emph{ResetQ} queue, which stores physical addresses that are initialized to all-0s content and 2) a 32-entry \emph{SetQ}, which stores physical addresses that are initialized to all-1s content.
The size of the SU is 256B. 

The \textit{fourth} component is the \emph{initialization queue} (\textbf{InitQ}). \tech{} uses this queue to record unused physical locations in PCM, such that they can be re-initialized methodically.
In terms of structure, InitQ is similar to the write queue but much simpler, in that each InitQ entry stores only address information and a single bit to indicate if the address is to be re-initialized with all-zeros or all-ones, while each write queue entry stores the write data as well the address.

We now describe the three key operations of \tech{}: 1) address translation, 2) overwritten content selection, and 3) re-initialization to all-0s and all-1s content.

\vspace{-10pt}
\subsubsection{Address Translation}\label{sec:translation}
\input{sections/address_translation}

\vspace{-5pt}
\subsubsection{Overwritten Content Selection}\label{sec:prioritization}
\input{sections/overwritten_content_selection}

\subsubsection{Re-Initialization of Unused Memory Locations}
\label{sec:scheduling}
\input{sections/scheduling}

\subsection{Other Considerations}
\label{sec:leveraging}
\input{sections/leveraging}

%% file: sections/address_translation.tex
\begin{figure}[h!]
	\centering
	\vspace{-5pt}
	\centerline{\includegraphics[width=0.99\columnwidth]{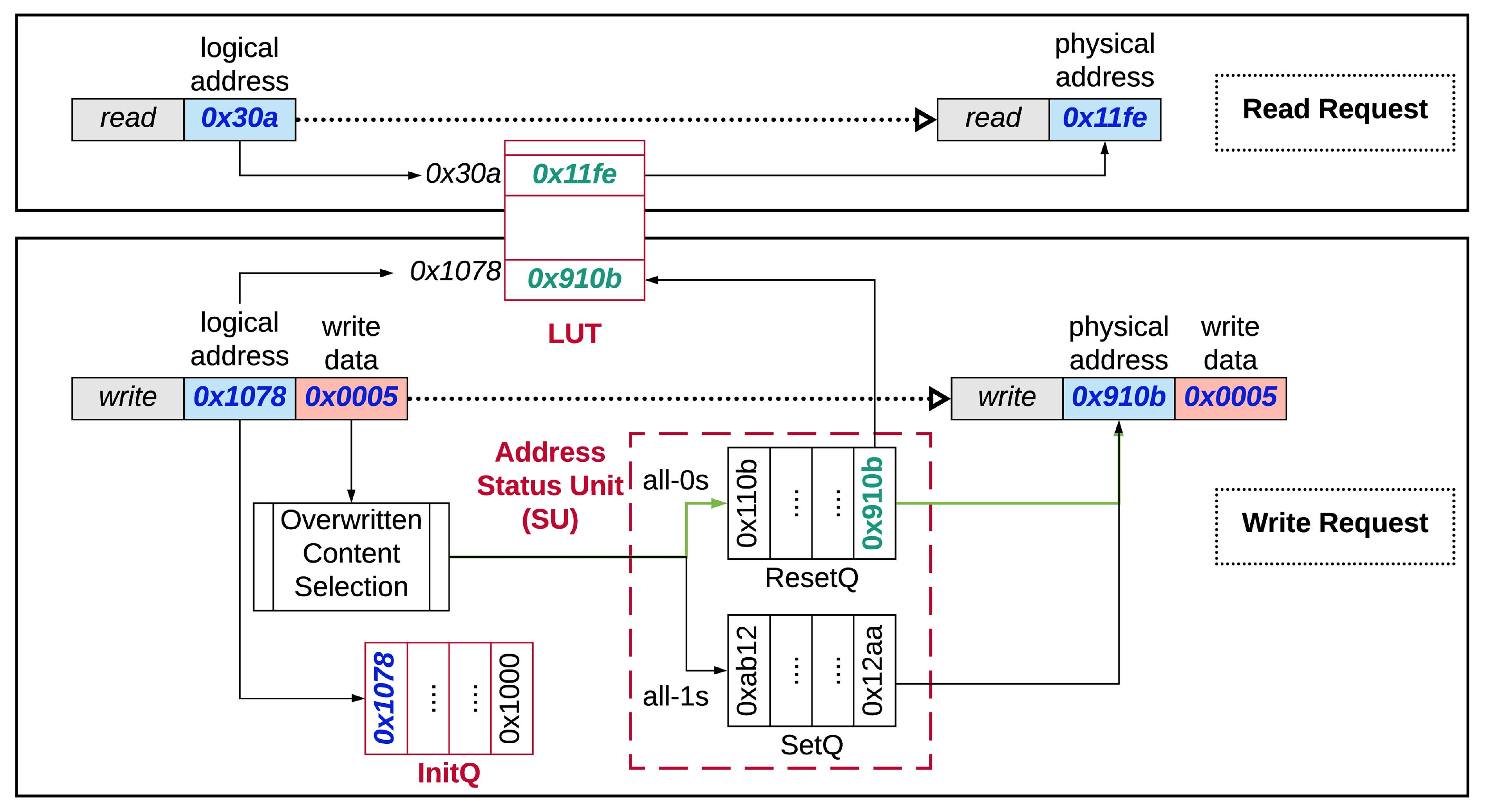}}
	\vspace{-10pt}
	\caption{Illustration of \tech{}'s address translation mechanism for servicing an example PCM read and write.}
	\vspace{-5pt}
	\label{fig:at_st_new}
\end{figure}

Figure~\ref{fig:at_st_new} shows the details of the address translation operation for an example read and an example write request, using the newly-introduced \tech{} components of the memory controller. To service a read request, the logical read address ({0x30a} in this example) is translated to the physical address {0x11fe} and scheduled to PCM. To service a write request, \tech{} first selects between overwriting all-0s and all-1s content based on the write data 0x0005 using the \emph{Overwritten Content Selection} unit (which we describe in Section \ref{sec:prioritization}). In this example, without loss of generality, we assume the overwritten content to be all-0s.
Therefore, \tech{} selects address 0x910b from the head of the ResetQ to redirect the write request to. The translation of logical address 0x1078 to physical address 0x910b is recorded in LUT. 
The newly-remapped address 0x1078 is also inserted in the InitQ to be re-initialized to all-0s or all-1s content (Section \ref{sec:scheduling}).

%% file: sections/overwritten_content_selection.tex
\begin{figure}[h!]
 	\centering
 	\centerline{\includegraphics[width=0.99\columnwidth]{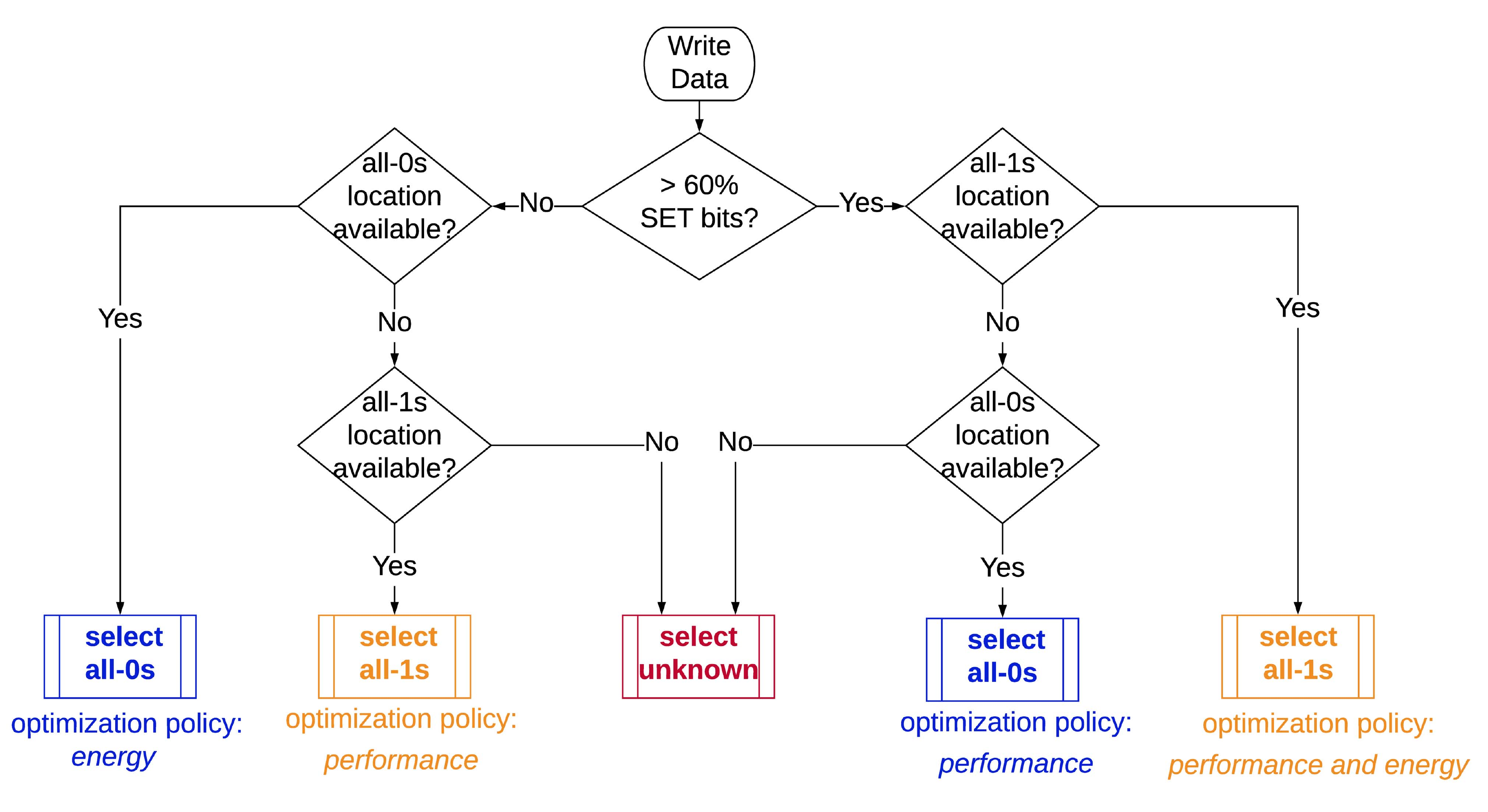}}
 	\vspace{-10pt}
 	\caption{Flowchart for overwritten content selection.}
    \vspace{-10pt}
 	\label{fig:flowchart}
\end{figure}

Figure \ref{fig:flowchart} illustrates \tech{}'s policy to select the overwritten content for every PCM write request. 
If the number of SET bits in the write data is more than 60\%, \tech{} checks the status table to see if there is all-0s or all-1s locations available in PCM.
If an all-1s location is available, \tech{} selects it as the overwritten content. This is because overwriting all-1s using only RESET operations is both more energy efficient and higher performance than overwriting any other content when the fraction of SET bits in the write data is more than 60\%. In this way, \tech{} optimizes both \emph{performance and energy}.
Otherwise, if an all-0s location is available, \tech{} selects it as the overwritten content. This is because overwriting all-0s using only SET operations has lower latency than overwriting unknown content. In this way, \tech{} optimizes \emph{performance}.  

Conversely, if the number of SET bits in the write data is less than 60\% and there is an all-0s location available, \tech{} selects it as the overwritten content. This is because overwriting all-0s using only SET operations is more energy efficient than overwriting any other content when the fraction of SET bits in the write data is less than 60\%. In this way,  \tech{} optimizes \emph{energy}. Otherwise, if an all-1s location is available, \tech{} selects it as the overwritten content. This is because overwriting all-1s using only RESET operations has lower latency than overwriting unknown content. 
 In this way, \tech{} optimizes \emph{performance}. 

\tech{} overwrites the existing content, which is unknown at the time of a PCM write only if there are no all-0s or all-1s locations available in PCM to service the write request. 


%% file: sections/scheduling.tex
As write requests are continuously re-mapped to different locations, DATACON re-initializes \emph{unused} memory locations with all-0s or all-1s so that future PCM write requests can be serviced using all-0s or all-1s locations.
To trigger such all-0s and all-1s re-initialization, we use \textit{thresholding}. 
Any time the number of entries in the ResetQ (which stores all-0s locations) or the SetQ (which stores all-1s locations) falls below the initialization threshold (\ineq{th_\text{init}}),
\tech{} issues a re-initialization request of a memory address stored in the InitQ.
However, the re-initialization request may not be serviced immediately.
\tech{} schedules re-initialization of memory locations off the critical path to minimize their interference with regular read and write accesses by exploiting a PCM bank's partition-level parallelism.
Re-initialization requests in \tech{} are serviced when
\begin{itemize}
	\item the read and the write queues are both empty, or
	\item the write queue is empty and a read request from the read queue is being serviced by a memory partition that is different from the one from which the re-initia\-lization request is to be serviced.
\end{itemize}

By using a threshold to trigger memory re-initial\-izations, \tech~ensures that the re-initialization overhead is reasonable. Furthermore, by sched\-uling these requests off the critical path, \tech's impact on system performance is also minimized (See Section \ref{sec:reinit}).

%% file: sections/leveraging.tex
We discuss two main issues that may effect \tech{}'s benefits in modern systems: data encryption and irregular memory accesses.
\subsubsection{\tech{} with Memory Encryption}\label{sec:mem_encryp}
Many persistent memory systems \cite{meza2013case,ren2015thynvm,zhao2014firm,lu2014loose,Volos2011,guerra2012software,marathe2017persistent,alshboul2018lazy} now include support for data encryption \cite{yang2005improving,GueronSP,HensonCSUR,chhabra2011nvmm,awad2017obfusmem,swami2016secret,saileshwar2018synergy,saileshwar2018morphable,ye2018osiris,awad2019triad}.
Enabling encryption inside the memory chip can lead to the following two issues. First, the write data going out of the CPU chip can change before overwriting the content at a memory location inside PCM. Second, 0s and 1s in the write data may be equally distributed for most write data.
\tech{} can be easily adapted to provide performance gain for such systems where encryption is enabled \textit{inside} the memory chip.
We observe that RESET operations are \textit{always} better than SET operations in terms of latency, while SET operations may be more energy efficient than RESET operations depending on the number of 0s and 1s in the write data. This follows directly from the two observations we made in Section \ref{sec:motivation}.

Therefore, when the write data content is unknown and/or equally distributed across 0s and 1s, we propose to configure DATA\-CON to prioritize only performance by overwriting all-1s using only RESET operations for every PCM write.
We analyze the performance improvement of this all-1s configuration of \tech{} in Section \ref{sec:results_perf_max}. We observe that by always overwriting all-1s, \tech{} can always improve performance 1) when the fraction of 0s and 1s in the write data are equal (as expected with encryption) and 2) \textit{independently} of the write data content. 

\subsubsection{\tech{} with Irregular Memory Accesses}
If a workload has many irregular memory accesses (e.g., as common in graph analytics \cite{ahn2015scalable,ahn2015pim,dogan2017accelerating,song2018graphr,nai2017graphpim} or pointer chasing workloads \cite{hsieh2016accelerating,boroumand2016lazypim,boroumand2018google,karlsson2000prefetching,collins2002pointer,kohout2001multi,wang2019stream,peled2019neural,hashemi2016accelerating,ebrahimi2009techniques}), the workload would likely generate many LUT misses. This is because PCM requests in such workloads tend to access many PCM partitions.
As \tech{} caches the address translation of only 2 PCM partitions inside LUT, most irregular accesses will miss in the LUT, generating extra PCM traffic.
To address this, we store the full AT in a separate dedicated PCM partition, which can be accessed in parallel with other partitions within a PCM bank using the bank's partition-level parallelism \cite{song2019enabling}. Doing so reduces the performance overhead of servicing the LUT misses. However, the energy overhead of LUT misses remains a concern. To address this, we store the AT in the eDRAM cache, which provides lower-latency and lower-energy access compared to PCM.

%% file: sections/evaluations.tex
\subsection{System Configuration}
To evaluate \tech{}, we designed a cycle-accurate DRAM-PCM hybrid memory simulator with the following:
\begin{itemize}
	\item A cycle-level in-house x86 multi-core simulator, whose front-end  is based on Pin~\cite{LukPin}. We configure this to simulate 8 out-of-order cores.
	\item A main memory simulator, closely matching the JEDEC Nonvolatile Dual In-line Memory Module (NVDIMM)-N/F/P Specifications~\cite{jedecnvdimm2017}. This simulator is composed of Ramulator~\cite{kim2016ramulator}, to simulate DRAM, and a cycle-level PCM simulator based on NVMain~\cite{poremba2015nvmain}. 
	\item Power and latency for DRAM and PCM are based on Intel/Micron's 3D Xpoint specification~\cite{bourzac2017has,villa2018pcm}. Energy is modeled for DRAM using DRAMPower~\cite{chandrasekar2012drampower} and for PCM using NVMain, with parameters from~\cite{villa2018pcm}.
\end{itemize}
Table \ref{tab:config} shows our simulation parameters. The memory controller implements a 16-entry read queue, a 16-entry write queue, and an 8-entry InitQ per each bank.

\begin{table}[h!]
	\centering
	\vspace{-5pt}
	\caption{Major Simulation Parameters.}
	\label{tab:config}
	\vspace{-5pt}
	{\fontsize{8}{10}\selectfont
		\begin{tabular}{lp{6cm}}
			\hline
			Processor & 8 cores per socket, 3.32 GHz, out-of-order\\
			\hline
			L1 cache & Private 32KB per core, 8-way\\
			\hline
			L2 cache & Private 512KB per core, 8-way\\
			\hline
			L3 cache & Shared 8MB, 16-way\\
			\hline
			eDRAM cache & Shared 64MB per socket, 16-way, on-chip\\
			\hline
			\multirow{5}{*}{Main memory} & 128GB PCM. \\
			& 4 channels, 4 ranks/channel, 8 banks/rank, 8 partitions/bank, 128 tiles/partition, 4096 rows/tile.\\
			& Memory interface = DDR4\\
			& Memory clock = 1066MHz\\
			& PCM Timings = See Table \ref{tab:min_max_latency}\\
			\hline
	\end{tabular}}
\end{table}

\vspace{-5pt}

\subsection{Evaluated Workloads}
We evaluate the following workloads.
\begin{itemize}
	\item 8 SPEC CPU2017 workloads~\cite{bucek2018spec}: bwaves, cactusBSSN, leela, mcf, omnetpp, parset, roms, and xalancbmk.
	\item 2 NAS Parallel workloads~\cite{bailey1991parallel}: bt and ua.
	\item 10 TensorFlow machine learning workloads~\cite{tfex}: mlp, cnn, gan, rnn, dcgan, bidirectional rnn (bi-rnn), autoencoder (autoenc), logistic regression (lr), random forest (rf), and word2vec.
\end{itemize}

\begin{figure}[h!]
	\centering
	\centerline{\includegraphics[width=0.99\columnwidth]{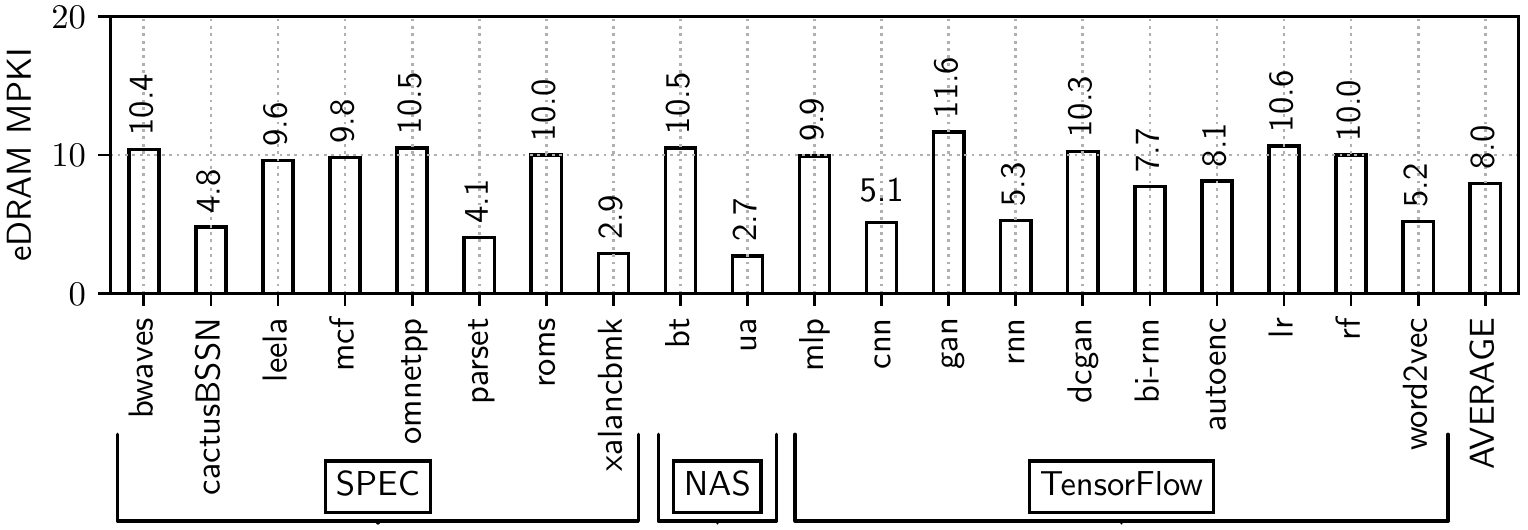}}
	\vspace{-10pt}
	\caption{eDRAM cache MPKI for the evaluated workloads.}
	\vspace{-10pt}
	\label{fig:mpki}
\end{figure}

All workloads are executed for 10 billion instructions. Figure~\ref{fig:mpki} plots the eDRAM Misses Per Kilo Instruction (MPKI) of the evaluated workloads.
For other workloads with low MPKI (not reported in Section~\ref{sec:results}), \tech{} neither significantly improves nor hurts performance and energy.

\begin{figure*}[b!]
	\centering
	\centerline{\includegraphics[width=1.95\columnwidth]{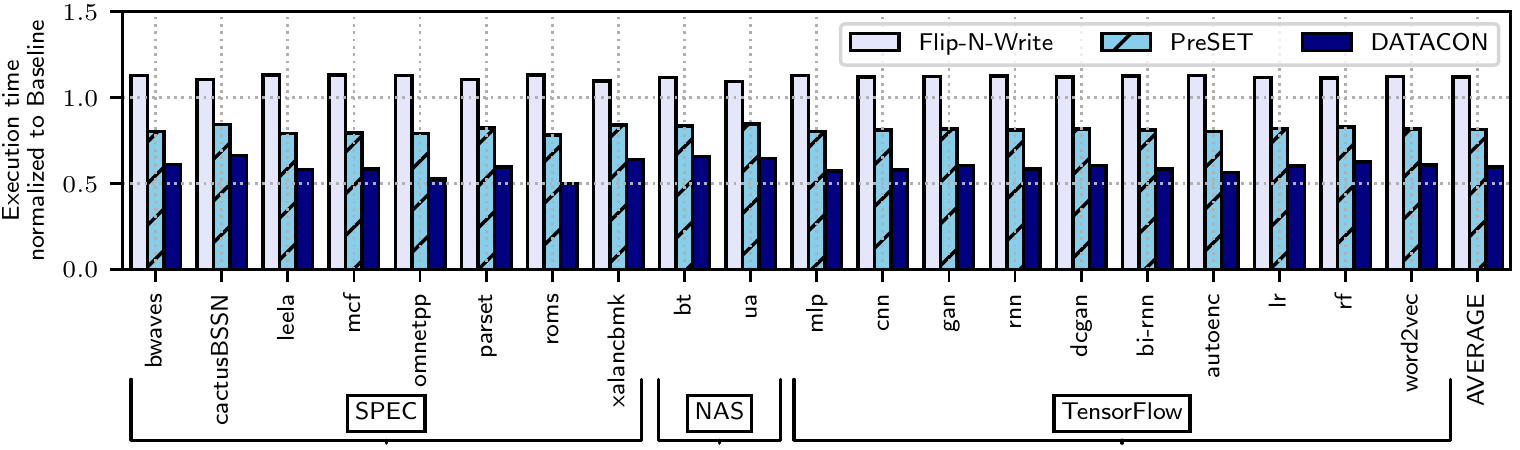}}
	\vspace{-10pt}
	\caption{Execution time normalized to Baseline for the evaluated workloads.}
	\vspace{-10pt}
	\label{fig:extime}
\end{figure*}

\begin{figure*}[b!]
	\centering
	\centerline{\includegraphics[width=1.95\columnwidth]{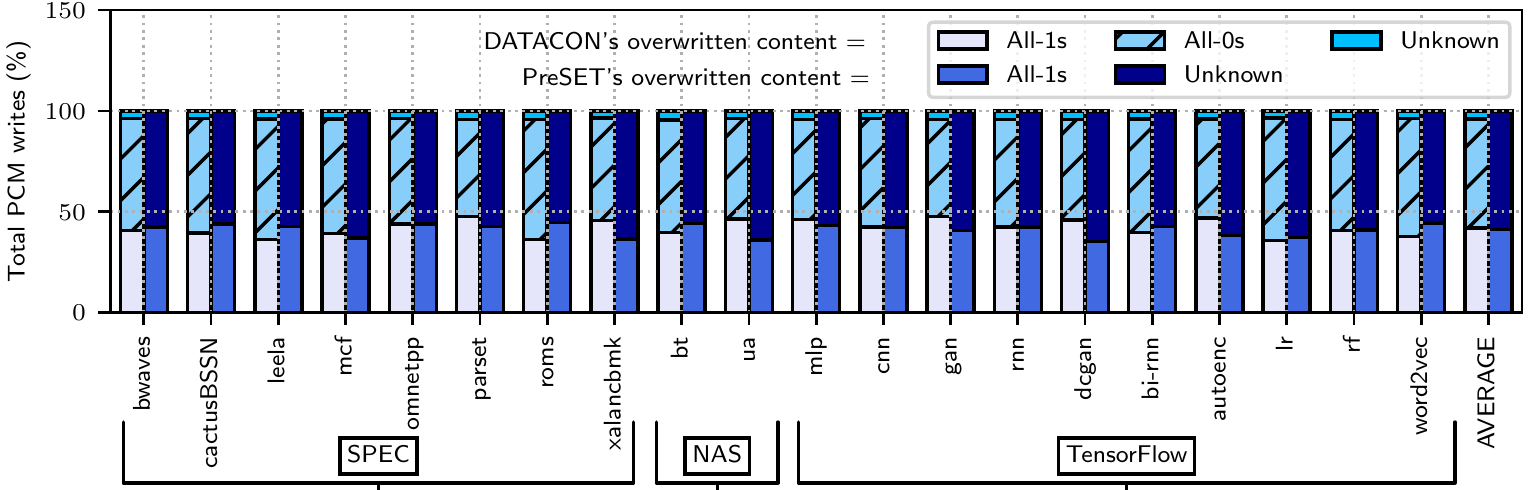}}
	\vspace{-10pt}
	\caption{Total PCM writes distributed into overwriting all-0s, all-1s, and unknown content in \tech{}, compared to overwriting all-1s and unknown content in PreSET \cite{QureshiISCA12}.}
	\vspace{-10pt}
	\label{fig:write_distribution}
\end{figure*}

\subsection{Evaluated Systems}
We evaluate the following systems.
\begin{itemize}
	\item \emph{Baseline}~\cite{LeeISCA2009} services PCM writes by overwriting \emph{unknown} content.
	\item \emph{PreSET}~\cite{QureshiISCA12}, which services PCM writes by \emph{always} overwriting all-1s. 
	\item \emph{Flip-N-Write}~\cite{cho2009flip} services PCM writes by \emph{finding out} the memory content using additional reads and programming only bits that are different from the write data. 
	\item \emph{\tech} redirects write requests to overwrite all-0s or all-1s content, depending on the write data. 
\end{itemize}

\subsection{Figures of Merit}
We report the following figures of merit in this work.


\begin{itemize}
	\item \emph{Access Latency:} The sum of queuing delay and service latency, averaged across all PCM accesses. 
	\item \emph{System Energy Consumption:} The sum of DRAM and PCM energy for all accesses.
	\item \emph{Execution Time:} The time to complete a workload.
\end{itemize}
\vspace{0.3pt}


%% file: sections/results.tex
\subsection{Overall System Performance}
\label{sec:perf}
Figure \ref{fig:extime} plots the execution time of each workload for each of the evaluated systems, normalized to Baseline. We make two main observations.

First, between PreSET and Flip-N-Write,  PreSET has higher performance. Average execution time of PreSET is 18\% lower than Baseline, while that of Flip-N-Write is 12\% higher than Baseline. PreSET improves performance over Baseline by programming PCM cells only in the RESET direction. Flip-N-Write, which finds out the memory content before overwriting it, has lower 
performance than Baseline due to the extra PCM read accompanying \textit{every} PCM write.

\begin{figure*}[b!]
	\centering
	\centerline{\includegraphics[width=1.95\columnwidth]{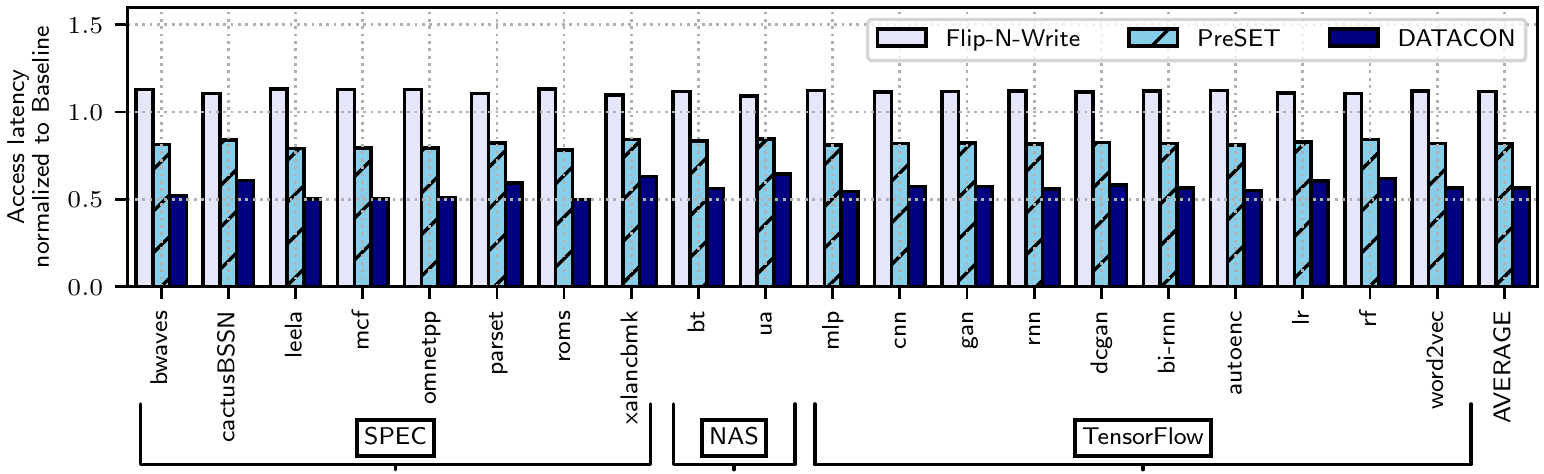}}
	\vspace{-10pt}
	\caption{Access latency normalized to Baseline for the evaluated workloads.}
	\label{fig:actime}
\end{figure*}

\begin{figure*}[b!]
	\centering
	\centerline{\includegraphics[width=1.95\columnwidth]{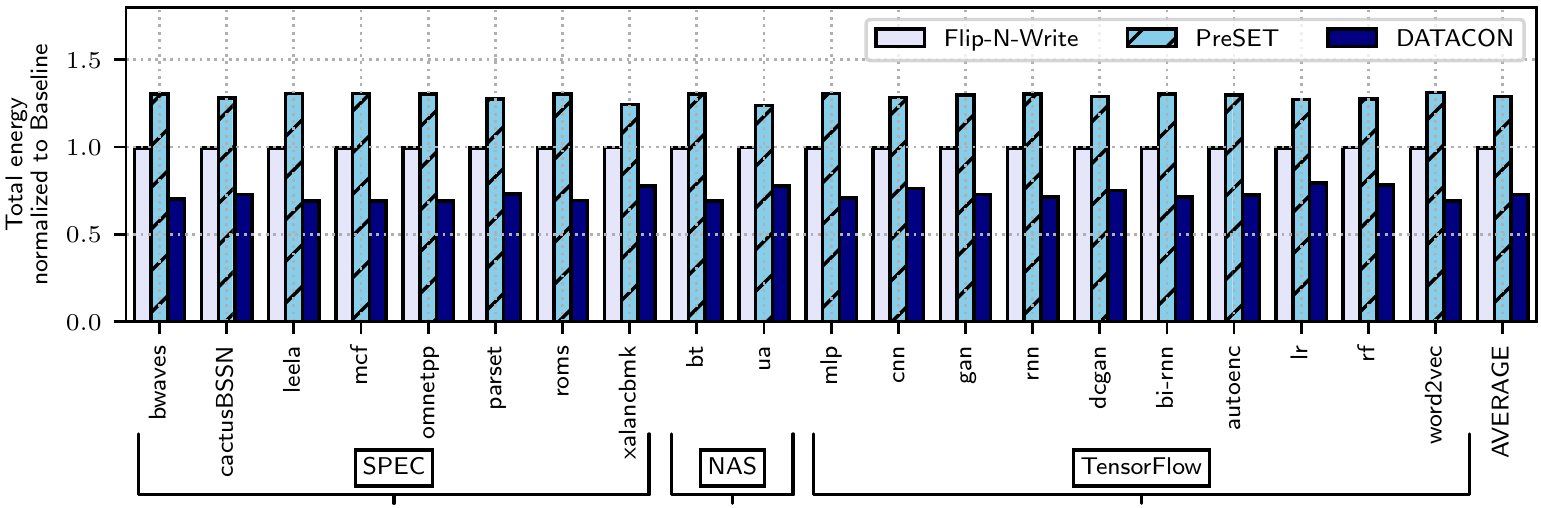}}
	\vspace{-10pt}
	\caption{Energy consumption normalized to Baseline for the evaluated workloads.}
	\vspace{-10pt}
	\label{fig:energy}
\end{figure*}

Second, \tech{} has the highest performance among all the evaluated techniques. Average execution time of DATACON is 40\% lower than Baseline, 47\% lower than Flip-N-Write, and \perfim\% lower than PreSET.
\tech{} outperforms PreSET because it greatly reduces the probability of overwriting unknown content on a write access. It does so in two ways. First, 
\tech{} overwrites all-0s as well, using only RESETs, when there is no all-1s location to overwrite, yet PreSET is limited to only overwriting all-1s, and it overwrites unknown content in the absence of opportunity to overwrite all-1s. 
As we show, overwriting all-0s provides 19\% latency reduction per access than overwriting unknown content (Section \ref{sec:quantifying_perf}).
Second, \tech{} methodically re-initializes unused memory locations to all-0s and all-1s content, whereas PreSET does so only opportunistically. We find that the probability of overwriting an all-0s or all-1s location is therefore much higher in \tech{} than in PreSET.  

To illustrate this, Figure~\ref{fig:write_distribution} plots the distribution of total PCM writes into the type of content they overwrite in \tech{} and PreSET. We observe that PreSET overwrites all-1s content for 41\% of PCM write accesses and unknown content for 59\%.
\tech{} overwrites all-0s content for 54\% of write accesses, all-1s content for 42\%, and unknown content for only 4\% of all PCM write accesses. \tech{} has higher performance and lower energy than PreSET because \tech{} greatly reduces the number of times unknown content is overwritten by also overwriting all-0s content in addition to all-1s content.

\vspace{-10pt}

\subsection{Access Latency}
\label{sec:effictive_rd_wr_latency}
\mcone{
Figure~\ref{fig:actime} plots average access latency of the workloads for each of our evaluated systems, normalized to Baseline. We make two main observations.
}

First, between PreSET and Flip-N-Write, PreSET has lower access latency than Flip-N-Write. Access latency of PreSET is 19\% lower than Baseline and that of Flip-N-Write is 10\% higher than Baseline. Flip-N-Write, which increases the queuing delay of a PCM write request to serve the extra read request needed to first know the memory content, has higher access latency than Baseline. PreSET, on the other hand, reduces access latency by programming PCM cells only in the SET direction.
Second, average access latency of \tech{} is 43\% lower than Baseline, 50\% lower than Flip-N-Write, and \aclatim\% lower than PreSET. 
The reason for \tech{}'s improvements are the same as explained in Section~\ref{sec:perf}.

\subsection{Energy Consumption}
\label{sec:energy}
\mcone{
Figure \ref{fig:energy} plots the total energy consumption of each of the workloads for each of our evaluated systems, normalized to Baseline. We make three main observations. 
}

First, the average energy consumption of Flip-N-Write is lower than Baseline by only 0.5\%. 
Although Flip-N-Write saves PCM write energy by minimizing the programming of PCM cells in a memory location based on the write data and the overwritten content, such energy saving is compensated by the energy increase due to the PCM read, which is issued before every write access to find out the overwritten data.


Second, between PreSET and Flip-N-Write, PreSET has an average 30\% higher energy consumption than Flip-N-Write.
High energy consumption in PreSET is due to three reasons. First, PreSET consumes energy to prepare a memory location with all-1s content before overwriting it. This preparation energy is higher than the energy consumed in reading the content as required by Flip-N-Write. Second, as PreSET does not analyze the write data, the energy consumption in PreSET can be high when there are many RESET operations to be performed to service a write. This is the case where the number of SET bits in the write data is low. As we have discussed in Observation 2 and reported in Figure \ref{fig:set_reset_distribution}, on average, only 33\% of PCM writes in the evaluated workloads has more than 60\% SET bits in the write data. Flip-N-Write has lower energy than PreSET in scenarios where the number of SET bits is low. This is because Flip-N-Write minimizes programming of PCM cells during a write by analyzing both the write data and the overwritten content. Finally, PreSET overwrites unknown content when there is no opportunity to prepare the memory content due to outstanding requests in the request queue. Overwriting unknown content increases energy consumption.

Third, although \tech{} incurs energy overhead for methodically re-initializing unused memory locations, the average energy consumption of \tech{} is still 27\% lower than Baseline, 26\% lower than Flip-N-Write, and \enim\% lower than PreSET.
This improvement is because \tech{} exploits the energy-latency trade-offs of SET and RESET operations of the PCM cells in selecting between using only SET or only RESET operations during a PCM write, unlike PreSET, which always uses \textit{only} SET operations, or Flip-N-Write, which uses \textit{both} SET and RESET operations during a PCM write. 
As explained in Section \ref{sec:perf}, \tech{} minimizes the probability of overwriting unknown content among the three mechanisms, which also reduces energy consumption.

\subsection{Re-Initialization Overhead}
\label{sec:reinit}
\begin{figure}[h!]
	\centering
	\vspace{-5pt}
	\centerline{\includegraphics[width=0.99\columnwidth]{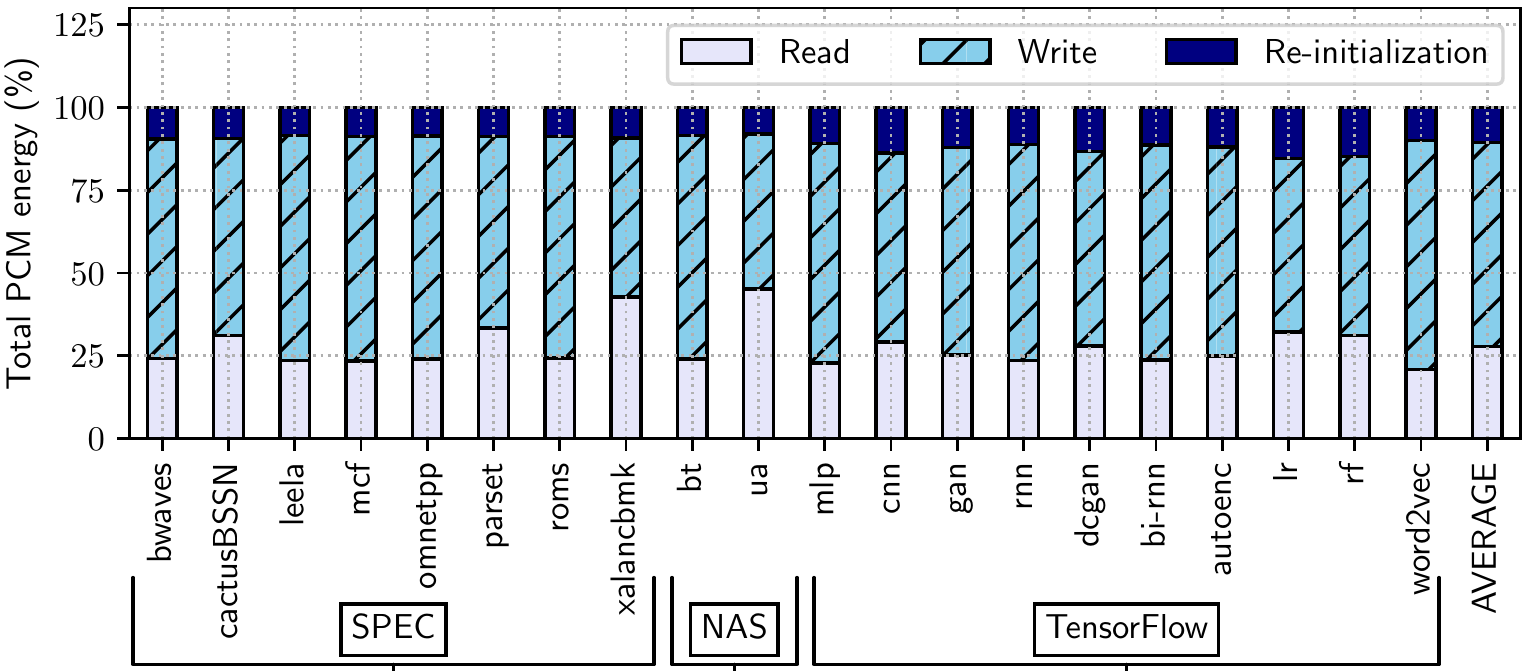}}
	\vspace{-10pt}
	\caption{\mcone{Total PCM energy distributed into energy to service reads, writes, and re-initialization requests.}}
	\vspace{-5pt}
	\label{fig:reinit}
\end{figure}

Figure \ref{fig:reinit} plots the total PCM energy distributed into energy to service reads, writes, and re-initialization requests in \tech{} for the evaluated workloads. 
We make the following two main observations from this data.

First, PCM reads and writes constitute, on average, 89\% of the total PCM energy.
The distribution of read and write energy within a workload depends on the relative proportion of PCM reads and writes in the workload. 
NAS\_ua, which has a higher proportion of PCM reads, has a higher fraction of total energy spent on reads
than NAS\_bt, which has higher proportion of PCM writes.
Second, re-initialization requests constitute, on average, 11\% of the total PCM energy.
This overhead is to service the re-initialization of unused memory locations in PCM, every time the available number of initialized all-0s or all-1s memory locations falls below the initialization threshold ($th_\text{init}$), which is set to 16.


\begin{figure*}[t]
	\centering
	\centerline{\includegraphics[width=1.95\columnwidth]{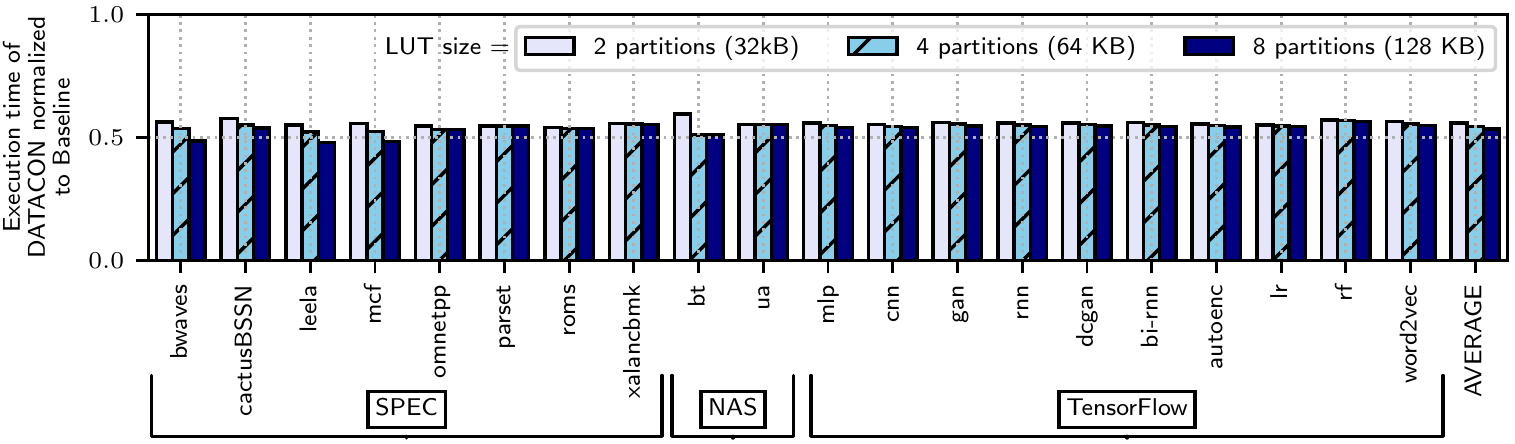}}
	\vspace{-10pt}
	\caption{{Execution time of \tech{} normalized to Baseline with three address translation cache sizes -- 2 partitions (32KB) (our default configuration), 4 partitions (64KB), and 8 partitions (128KB).}}
	\vspace{-10pt}
	\label{fig:caching}
\end{figure*}

\begin{figure*}[t!]
	\centering
	\centerline{\includegraphics[width=1.95\columnwidth]{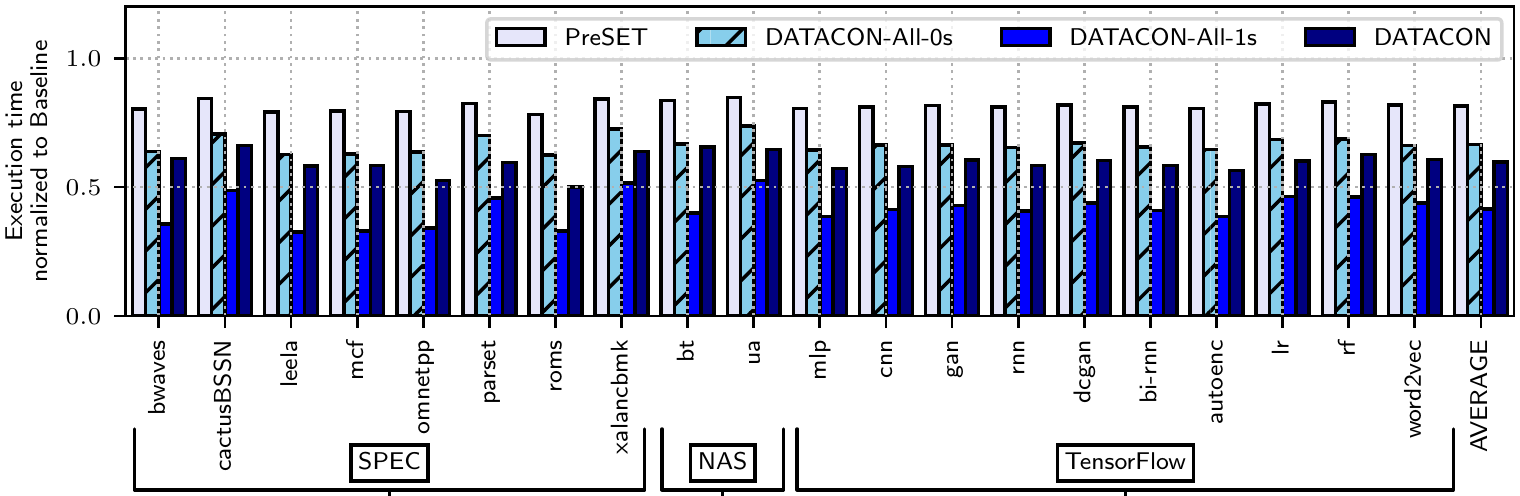}}
	\vspace{-10pt}
	\caption{{Execution time of \tech{}-all-0s and \tech{}-all-1s normalized to Baseline for the evaluated workloads.}}
	\vspace{-10pt}
	\label{fig:perf_zeros_ones}
\end{figure*}

\begin{figure*}[b!]
	\centering
	\centerline{\includegraphics[width=1.95\columnwidth]{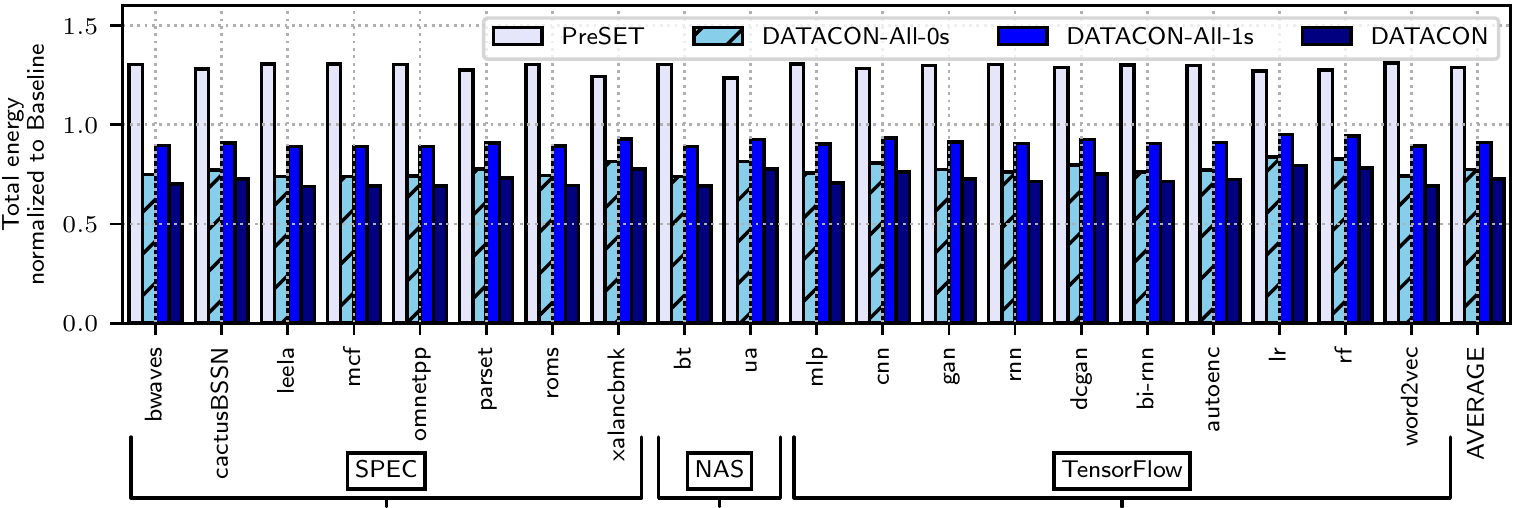}}
	\vspace{-10pt}
	\caption{{Energy consumption of \tech{}-all-0s and \tech{}-all-1s normalized to Baseline for the evaluated workloads.}}
	\vspace{-10pt}
	\label{fig:energy_zeros_ones}
\end{figure*}

\subsection{LUT Sizing}
\label{sec:caching}
Figure \ref{fig:caching} plots \tech{}'s execution time normalized to Baseline for each of the evaluated workloads 
when the LUT size is increased from 2 partitions (our default configuration) to 8 partitions. 
We make two main observations.

First, when the LUT size is increased, the number of LUT misses reduces and thus performance increases. With only 4 and 8 recently-used PCM partitions cached in the LUT, the average execution time of \tech{} is respectively, only 3\% and 5\% lower than \tech{} with the default configuration. 
Second, for most workloads, caching the address translation of 2 recently-used PCM partitions is sufficient to provide high performance and there is only marginal performance improvement from caching more partitions. This is because workloads exhibit a high degree of partition-level spatial locality in accesses, which \tech{} exploits to its benefit in keeping the address translation overhead reasonable.

\vspace{-10pt}

\subsection{Overwriting Only All-0s or All-1s Content in \tech{}}
\label{sec:results_perf_max}
We configure \tech{} to overwrite only all-0s and only all-1s content for every single PCM write, to evaluate the benefits of having the ability to decide between all-0s and all-1s depending on overwritten content. We call these two modes of \tech{} as \emph{\tech{}-all-0s} and \emph{\tech{}-all-1s}, respectively. 
\tech{}-all-1s could also be a good mechanism to employ when memory content is encrypted inside the memory chip, as explained in Section~\ref{sec:mem_encryp}.

Figure \ref{fig:perf_zeros_ones} plots the execution time of \tech{}, DATA\-CON-all-0s, and \tech{}-all-1s normalized to Baseline for the evaluated workloads. We make the following three main observations.

First, between \tech{}-all-0s and DATA\-CON-all-1s, DATA\-CON-all-1s has higher performance. The average execution time of \tech{}-all-1s is 58.5\% lower than Baseline and that of \tech{}-all-0s is 34\% lower than Baseline. As discussed in Section~\ref{sec:quantifying_perf}, to overwrite all-1s content using \tech{}-all-1s, PCM cells are programmed in the RESET direction, which has lower latency than programming the PCM cells in the SET direction as required by \tech{}-all-0s to overwrite all-0s content.

Second, although both PreSET and \tech{}-all-1s overwrite all-1s at every PCM write, the performance of DATA\-CON-all-1s is, on average, 50\% higher than PreSET. This is because PreSET 
requires the memory location to be initialized to all-1s content first before overwriting it during a write. If there are outstanding requests in the request queue, PreSET skips the initialization step and proceeds to overwrite unknown content, thereby hurting performance. \tech{}-all-1s, on the other hand, overwrites any available all-1s location in PCM by translating the write address. It also methodically re-initializes locations to all-1s to maximize the probability of overwriting them when servicing write requests. Therefore, \tech{}-all-1s overwrites all-1s content for more PCM writes (on average, 2.3x more) than PreSET does, thereby achieving higher performance.

Third, \tech{}, which selects between overwriting all-0s and all-1s based on the write data and the energy-performance trade-offs, has higher performance than DATA\-CON-all-0s (average 10\% lower execution time) and lower performance than \tech{}-all-1s (average 46\% higher execution time). Despite this relatively lower performance of \tech{} compared to \tech{}-all-1s, \tech{} outperforms PreSET by achieving \perfim\% lower average execution.

Figure~\ref{fig:energy_zeros_ones} plots the total energy consumption of the four mechanisms for the evaluated workloads. We make two main observations. First, \tech{}-all-1s leads to higher energy consumption than DATA\-CON-all-0s because RESET operations used by \tech{}-all-1s require higher energy than SET operations used by \tech{}-all-0s. Second, \tech{} has lower energy consumption than the other mechanisms because \tech{} selects between SET and RESET operations by effectively considering the fraction of 1s versus 0s in the write data and the associated energy-performance trade-offs of SET and RESET operations (as explained in Section~\ref{sec:motivation}).

\subsection{Performance and Energy Trade-off}
\label{sec:tradeoffs}
\begin{figure}[h!]
	\centering
	\vspace{-5pt}
	\centerline{\includegraphics[width=0.99\columnwidth]{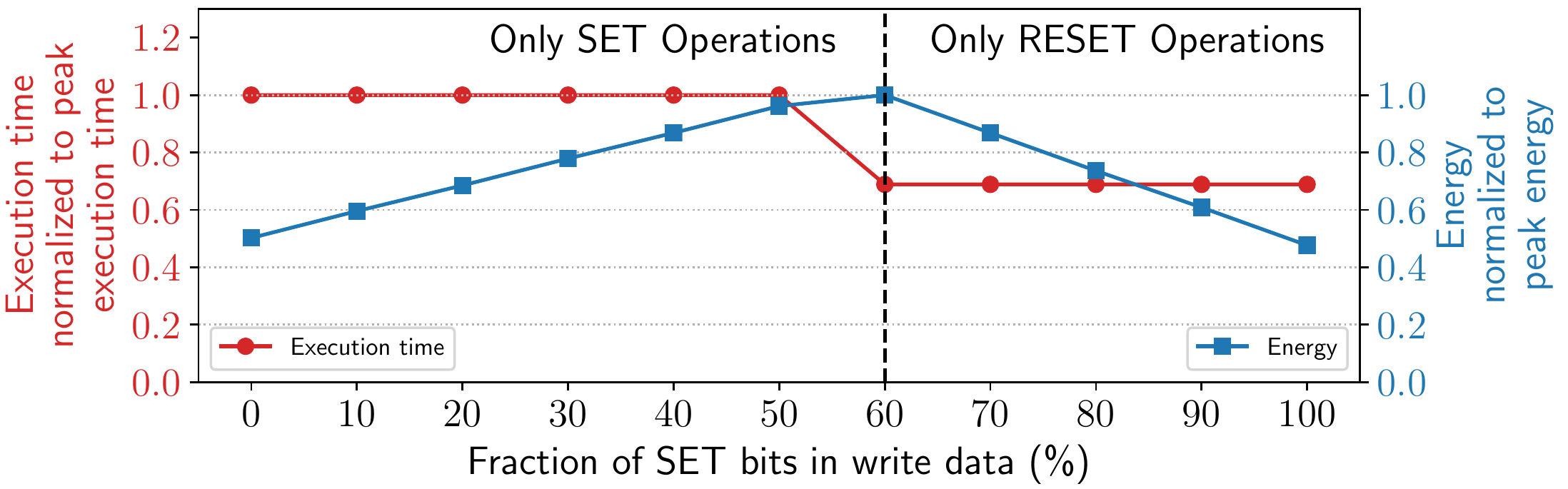}}
	\vspace{-10pt}
	\caption{Execution time and energy of \tech{} normalized to their peak values with increasing fraction of SET bits in the write data of a microbenchmark.}
	\vspace{-5pt}
	\label{fig:tradeoff}
\end{figure}

To analyze the performance and energy trade-off evaluation performed by \tech{} in response to the write data, we designed a microbenchmark, where the \textit{same} write data is used for \textit{every} PCM write access. We ran the microbenchmark repeatedly while increasing the fraction of set bits in the write data every time.
Figure \ref{fig:tradeoff} plots the execution time and energy of \tech{} normalized to peak points in each curve when increasing the fraction of set bits from 0\% (write data is all-0s) to 100\% (write data is all-1s).
We make the following two main observations.

First, as the fraction of set bits in write data is increased, energy consumption increases, reaching a peak at around 60\%, and then decreases. This is due to the energy difference of SET and RESET operations, which we have analyzed in Section~\ref{sec:quantifying_perf} and illustrated in our Observation 1 in Figure~\ref{fig:set_reset_energy}. \tech{} lowers the energy consumption by switching from only SET operations (overwriting all-0s) to only RESET operations (overwriting all-1s) as the fraction of SET bits in the write data exceeds 60\%.
Second, as the RESET operation in PCM has lower latency than the SET operation (see Table~\ref{tab:min_max_latency}), switching to using only RESET operations also leads to lower execution time, i.e., higher performance.

\subsection{Effect on PCM Lifetime}
\begin{figure}[h!]
	\centering
	\vspace{-5pt}
	\centerline{\includegraphics[width=1.01\columnwidth]{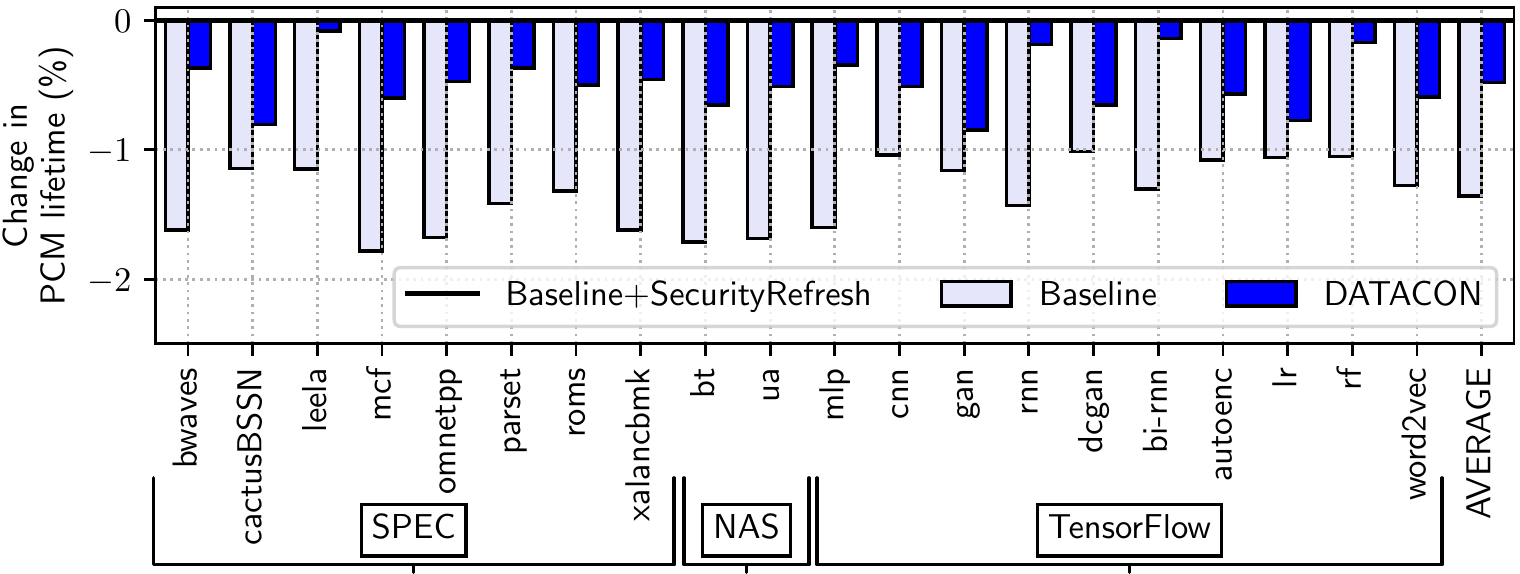}}
	\vspace{-10pt}
	\caption{\mcone{Effect of DATACON on PCM Lifetime relative to Baseline with SecurityRefresh \cite{SeongSecurityISCA2010}.}}
	\vspace{-10pt}
	\label{fig:endurance}
\end{figure}

We compare the PCM lifetime (in years) obtained with DATACON against Baseline and Baseline+SecurityRefresh, which is our Baseline extended with a dynamic address translation scheme called SecurityRefresh~\cite{SeongSecurityISCA2010}, a state-of-the-art wear-leveling mechanism for PCM. We assume a pessimistic PCM cell endurance of $10^{7}$ writes. Figure~\ref{fig:endurance} shows the impact of \tech~on PCM lifetime for the evaluated workloads, normalized to the lifetime obtained using Baseline+Security\-Refresh. We also plot the lifetime with Baseline (without SecurityRefresh). We make two main observations.

First, 
Baseline has an average 1.35\% lower lifetime than Baseline+Secu\-rityRefresh. This is because the dynamic address translation mechanism in Baseline+SecurityRefresh distributes the PCM write requests of a workload to other locations, balancing the wear-out of PCM cells. Baseline, on the other hand, writes to only the memory locations requested by the program, inducing higher wear-out in the PCM cells in those requested locations.\footnote{SecurityRefresh's lifetime improvement over Baseline is lower than that reported in~\cite{SeongSecurityISCA2010} because of the eDRAM write cache used in our system, which mitigates PCM's endurance problem by greatly reducing the number of write accesses to PCM.}
Second, the PCM lifetime with \tech~is slightly lower than Baseline with SecurityRefresh by, on average, 0.48\%, despite both techniques performing dynamic address translation when servicing PCM writes.\footnote{\tech{} has 25\% higher performance and 40\% lower energy than Baseline+SecurityRefresh.} This is due to the different optimization objectives of the two techniques. When selecting a physical address for a write request, \tech~searches for a memory location that contains the best data pattern for the write data that would result in the lowest latency and energy. 
For Baseline with SecurityRefresh, the objective is to redirect a write request to a memory location  that has the \textit{least wear-out}, but the content is unknown at the time of the write.
We observe that the impact of PreSET and Flip-N-Write on PCM lifetime are both similar to Baseline \textit{without} SecurityRefresh. This is because both PreSET and Flip-N-Write do not use address translation, which helps to level the wear-out in PCM. DATACON achieves slightly better PCM lifetime than Baseline, Flip-N-Write, and PreSET. We believe DATACON can be combined with SecurityRefresh in the future to achieve even smaller impact on PCM lifetime.

%% file: sections/related.tex
\mcone{
To our knowledge, this is the first work that improves both performance and energy consumption in PCM by analyzing only the data to be written,
without having to first read the overwritten content.
}


\subsection{Related Concepts in Flash Memory}
The Erase and Program operations in Flash memory are equivalent to the SET and RESET operations in PCM with similar asymmetries in energy and latency.
To facilitate erase-before-write in Flash memory, logical-to-physical address translations are performed to hide the long latency of erase operations~\cite{gupta2009dftl,ma2014survey,kim2002space}.
Invalidated physical locations are erased periodically
\cite{chang2004real,cai2017error}.

\tech~is different from the erase-before-write in Flash memory for three reasons.
First, \tech~selects between overwriting all-0s and all-1s based on energy-latency trade-offs, whereas writing in Flash memory is \emph{always} performed on erased content (i.e., on all-1s).
Second, the granularity of read, write and re-initialization operations are the same in \tech, whereas the granularity of an erase operation in Flash memory is much \emph{larger} than the granularity of a write operation, which leads to different performance and energy trade-offs.
Third, \tech~is an \emph{optional} operation and tries to maximize overwriting all-0s or all-1s, whereas an erase is a \emph{required} operation before writing a block in Flash memory.
Inspired from the erase-before-write in Flash memory, block erase is proposed for PCM~\cite{lam2010block}.



\subsection{Writeback Optimizations}
Line-level writeback is proposed in \cite{QureshiISCA09,LeeISCA2009,pourshirazi2018wall}, which tracks the lines in a memory page that are dirty and selectively writes only those lines to PCM when a DRAM/eDRAM cache line is evicted.
\tech{} implements the line-level writeback policy \emph{only} to overwrite unknown content. 
Dynamic write consolidation is proposed in~\cite{seshadri2014dirty,xia2014dwc,StuecheliISCA,wang2013wade,lee2010dram} to consolidate multiple writes targeting the same row into one write operation. 
Write activity reduction is proposed in~\cite{hu2013write,huang2011register}, which uses registers for servicing costly PCM writes.
Multi-stage PCM write is proposed in ~\cite{qureshi2010improving,qureshi2010morphable,yue2013accelerating,zhang2016mellow} to improve PCM performance by completing a long-latency write access to PCM in several steps, with each step scheduled off the critical path of read accesses.
\tech~is complementary to all these approaches.


\subsection{Read Before Write}
Read before write is proposed to reduce the number of SET operations when servicing PCM write accesses.
One example is the {Flip-N-Write}~\cite{cho2009flip}, which compares the overwritten content with the \emph{write data} to decide if writing an inverted version of the data would result in a lower number of SET operations than writing the requested data itself.
Other works propose variants of this approach~\cite{chen2011energy,yang2007low,mirhoseini2012coding,fang2012softpcm,jacobvitz2013coset}, where the write data is differentially encoded with the overwritten content. The differentially-encoded data is used to overwrite the content.
All these techniques convert a PCM write operation into a PCM read, followed by a PCM write. We compare \tech{} to Flip-N-Write and find that DATACON performs significantly better (Section \ref{sec:results}).


\subsection{Related Works in Multi-Level PCM}
In recent PCM devices, PCM cells are used to store multiple bits per cell (referred to as multilevel cell or MLC).
MLC PCM offers greater capacity per unit area at the cost of asymmetric latency and energy in accessing the bits in a cell.
Yoon et al. propose an architectural technique for data placement in MLC PCM~\cite{yoon2014efficient}, exploiting latency and energy asymmetries.
Qureshi et al.  propose a morphable PCM system~\cite{qureshi2010morphable}, dynamically adapting between high-density and high-latency MLC PCM and low-density and low-latency single-level cell PCM.
Qureshi et al. propose write cancellation and pausing to allow reads to be serviced faster by interrupting long write operations in PCM~\cite{qureshi2010improving}.
Jiang et al. propose write truncation~\cite{jiang2012fpb}, where a write operation is truncated to allow read operations, compensating for the loss in data integrity with stronger ECC.
These techniques are also complementary to and can be combined with \tech{}.

%% file: sections/conclusions.tex
We introduce \tech{}, a simple and effective mechanism to reduce the latency and energy of write operations in Phase Change Memory (PCM).
\tech{} is based on the key observation that overwriting \emph{unknown} memory content can incur significantly higher latency and energy compared to overwriting \emph{known} all-zeros or all-ones data pattern. 

Based on this observation, \tech~redirects PCM write requests to overwrite memory locations containing all-zeros or all-ones, and overwrites unknown content only when it is absolutely necessary to do so. The key idea is to estimate how much a PCM write access would benefit from a particular \emph{known} content (i.e., all zeros or all ones) by comprehensively considering the fraction of set bits in the write data and the energy-latency trade-offs for SET and RESET operations in PCM.
Based on this estimate, \tech{} translates the logical write address to a physical address in PCM containing the best overwritten content and records this address translation in a table to service future requests. \tech{} exploits the data access locality in workloads to cache a small portion of address translation table inside the memory controller, lowering the address translation latency. \tech{} methodically re-initializes \emph{unused} memory locations to minimize the probability of overwriting unknown content, which improves both performance and energy efficiency.
Results of our experiments with workloads from SPEC CPU2017, NAS Parallel Benchmarks, and state-of-the-art machine learning applications show that \tech~improves the effective access latency by \aclatim\%, system performance by \perfim\%, and overall system energy consumption by \enim\% compared to the best of previously-proposed state-of-the-art techniques. 

We \textbf{conclude} that DATACON is a simple yet efficient technique to reduce the latency and energy of PCM writes, which significantly improves performance and energy efficiency of modern memory systems.
